\documentclass[aps,pra,showpacs,twoside,10pt,floatfix,nofootinbib,longbibliography,twocolumn]{revtex4-2}
\usepackage[colorlinks=true, citecolor=red, urlcolor=blue ]{hyperref}
\usepackage{amssymb,amsfonts,amsmath,bm,amsthm,mathtools,braket,geometry,xcolor,times,comment}
\usepackage[normalem]{ulem}
\newcommand{\bracket}[2]{\langle{#1}|{#2}\rangle}
\definecolor{boldcolor}{HTML}{703AA6}
\newcommand{\proposition}[1]{\vspace{0.1cm}\noindent\textbf{\textcolor{boldcolor}{$\blacksquare$ Proposition #1.}}}

\begin{document}

\title{Scalable and deterministic Greenberger-Horne-Zeilinger state generation via graph states-assisted measurements}
\author{Harikrishnan K J and Amit Kumar Pal}
\affiliation{Department of Physics, Indian Institute of Technology Palakkad, Palakkad 678 623, India}
\date{\today}

\begin{abstract}
We propose a scalable and deterministic protocol for growing large multi-qubit states starting from two-qubit non-maximally entangled pure states, where the bipartite entanglement in the resultant state is higher than the maximum of the available entangled qubit-pairs. This is achieved via a truncation of the Hilbert space corresponding to a subsystem of qubits to a space that hosts a single qubit, brought about by a multi-qubit measurement assisted by the graph basis. We prove its equivalence to a repetitive two-qubit measurement-based protocol, and demonstrate realization of the required two-qubit measurement via  a two-qubit parity measurement, thereby establishing the implementability of the protocol. We derive lower and upper bounds of the bipartite entanglement concentrated after a given number of rounds of measurements, where the entanglement of the available qubit-pairs are not-necessarily equal. We further discuss the effect of possible imperfections that may arise in the protocol, and its robustness towards such imperfections. We demonstrate the usefulness of our proposal by applying it to create generalized GHZ states on arbitrary number of qubits, thereby underlining the possibility of creating maximally entangled qubit pairs via qubit-local projection measurements.
\end{abstract}

\maketitle

\section{Introduction}
\label{sec:introduction}

As we witness the advancement of quantum  technologies at a faster pace, the idea of quantum networks~\cite{Perseguers2013,Kozlowski2019,Azuma2021,wei2022} have gathered a lot of momentum, fueled specifically by envisioned applications in secured as well as unsecured quantum communication~\cite{Gisin2002,Gisin2007,sende2011}, developing quantum internet~\cite{Kimble2008,Wehner2018,Kumar2025}, improving communication complexity~\cite{Guerin2016}, clock synchronization~\cite{Komar2014}, distributed quantum computation~\cite{Cirac1999} and sensing~\cite{Guo2020}. Various quantum protocols, including establishing viable quantum communication~\cite{Bennett1992,Bennett1993,Ekert1991,Mattle1996,Bouwmeester1997} require sharing entanglement~\cite{plenio1998,Horodecki2009}, used as resource for running these quantum protocols, among distantly located parties that are part of the established quantum network. This calls for practical strategies of generating quantum states having specifically tailored entanglement properties over all, or a selected subset of parties constituting the quantum network. 

Protocols for generating an entangled state with desired entanglement properties on an arbitrary number of parties generally rely on merging smaller entangled \emph{resource states} through entangling operations and measurements. These protocols can be either \emph{probabilistic}, where there can be a chance of failure of the enterprise, or \emph{deterministic}, when the desired entangled state is created with certainty. The former avenue often relies on merging smaller entangled states via performing suitable measurements on selected subsets of qubits with a subsequent post-selection of a convenient measurement outcome aiding the purpose~\cite{raussendorf2001,Sende2005,Walther2005,Acin2007,briegel2009,Perseguers2008,Perseguers2010,Cavalcanti2011,Banerjee2020a,Halder2021,OhSooMin2024}, including the swapping and repeater-based constructions~\cite{Zukowski1993,Briegel1998,Dur1999a,Wallnofer2016,Su2016}, while in the deterministic methods, the available smaller entangled states are fused together via applications of nonlocal unitary operations on chosen qubits~\cite{Ozaydin2014,Halder2022}. In both cases, entanglement in the generated multi-qubit states follow the fundamental restrictions on state convertibility~\cite{Morikoshi2001,Lo2001,Nielsen1999,Vidal1999,Jonathan1999}, entanglement concentration~\cite{Bennett1996,Hardy1999,Zhao2001,Yamamoto2001,Lo2001,Bose1999,Morikoshi2001,Morikoshi2000,Gu2006,Zukowski1993,Zukowski1995,sende2003a,Santra2021,Zhao2013} and purification~\cite{Bennett1996b,Horodecki1997,Murao1998,Dur1999,Horodecki1999,Pan2003,Zhao2003,Yamamoto2003} schemes.  

We go beyond the two distinct prescriptions for growing large entangled networks, and ask the following questions.
\begin{enumerate}
    \item[(a)]  \emph{Is it possible to design a measurement-based protocol for fusing smaller entangled states into a larger one, which is inherently probabilistic in nature, and yet operates as a deterministic protocol as long as the entanglement properties of the generates states are concerned?}
    \item[(b)] \emph{Can such a protocol be tuned to create a multi-qubit entangled state with a bipartite entanglement higher than the same present in the smaller entangled states used as constituents?}
    \item[(c)] \emph{If such a protocol exists, can it be employed to create important multi-qubit states useful for quantum information processing?}
\end{enumerate}

In this paper, we answer all of these three questions affirmatively. We demonstrate that starting from $N$ entangled qubit-pairs, one can use a specially designed $N$-qubit measurement to generate $(N+1)$-qubit entangled states having higher bipartite entanglement in a specific $1:\text{rest}$ bipartition compared to the entanglement present in the two-qubit constituent states. We further show that the generated $(N+1)$-qubit states are equally probable, and are connected to each other via unitary operators local to individual qubits, thereby making the protocol deterministic with respect to  the entanglement properties of the generated states. This  allows one to avoid the post-selection of the measurement outcome, and work with any of the outcome states with identical entanglement properties. We also employ the protocol to generate the Greenberger–Horne–Zeilinger (GHZ) state~\cite{greenberger1989} -- arguably one of the most important multipartite quantum states -- which is useful in a wide variety of contexts, including multi-party quantum secret sharing~\cite{Hillery1999,Gottesman2000}, measurement-based quantum computation~\cite{raussendorf2001,briegel2009}, quantum error corrections~\cite{gottesman2010}, enhancing precision of quantum measurements~\cite{Giovannetti2004}, and purifying Bell states on arbitrary qubit-pairs via strategic single-qubit measurements~\cite{divincenzo1998,popp2005,hein2006}. Recent advancements in generating large GHZ state in photonic~\cite{photonic_ghz_Tsujimoto2018,photonic_ghz_cao_2024,photonic_ghz_Melkozerov_2026} and atomic~\cite{atomic_ghz_Zhao_2021} systems also highlights the relevance of our work.

Specifically, we use an $N$-qubit measurement assisted by the graph-state basis that truncates the $N$-qubit Hilbert space to a single-qubit Hilbert space.  We prove that application of the measurement on the qubits labeled by $B_i$ in $N$ non-maximally entanglement qubit-pairs $A_iB_i (i=1,2,\cdots,N)$ results in a post-measured  ensemble of $2^N$ $(N+1)$-qubit states on the qubits $BA_1A_2\cdots A_N$, where $B$ corresponds to the single-qubit Hilbert space obtained by truncating the $2^N$-dimensional Hilbert space of the qubits $B_1B_2\cdots B_N$ (see Fig.~\ref{fig:schematic}(a)). We show that each of these $(N+1)$-qubit states occurs with a probability $2^{-N}$, and are connected via unitary operators $U=U_{B}\bigotimes_{i=1}^NU_{A_i}$, thereby making the protocol deterministic with respect to entanglement properties of these states. We further prove that each of these $(N+1)$-qubit states have a bipartite entanglement over the partition $B:A_1\cdots A_N$ higher than the maximum of the bipartite entanglement present in the qubit pairs $A_iB_i$, which we refer to as the \emph{profitable concentration of entanglement} (PCE), and derive the lower and upper bounds of the concentrated bipartite entanglement in terms of the entanglement of the initial states.

Given the difficulty in implementing the $N$-qubit measurement in the laboratory, we develop an iterative variant of the protocol where two-qubit measurements are used instead of the $N$-qubit measurements. We establish the equivalence of these two protocols in terms of the entanglement properties of the multi-qubit outcome states, and prove that PCE is always achievable in a scalable fashion even in scenarios where the constituent entangled qubit-pairs have different entanglement. We derive the bounds of the concentrated entanglement for the modified scalable protocol, and  analyze the resultant $(N+1)$-qubit states using their multipartite entanglement content and monogamy properties. 

We further extend our analysis to the situations where the two-qubit measurements are noisy with the imperfections arising out of the single-qubit measurement involved in implementing the two-qubit measurement, and demonstrate that repeated application of the faulty single-qubit measurement can overcome the adversity due to noise. On the other hand, when noise enters via imperfectly prepared initial two-qubit states, our calculations reveal that a PCE is possible for all noise strengths when the initial two-qubit states are undergoing single-qubit phase flip noise. In contrast, in the case of the bit-flip and depolarizing noise, PCE is achievable only when the noise strength is below a threshold. We employ our proposal for network-growth to generate GHZ states on arbitrary qubits, and determine the detailed circuit description of the protocol.     

The rest of the paper is organized as follows. In Sec.~\ref{sec:N-qubit_measurement}, we introduce the graph states-assisted joint $N$-qubit measurement scheme, and establish the measurement-based yet deterministic protocol for preparing multi-qubit entangled states with profitable entanglement concentration. The protocol is modified via a repeated two-qubit measurement scheme in Sec.~\ref{sec:repeated_two_qubit_measurement}, while the possible effects of noise on the protocol is discussed in Sec.~\ref{sec:imperfections}. In Sec.~\ref{sec:GHZ_preperation}, we demonstrate the application of the protocol for preparing GHZ states on arbitrary number of qubits. Sec.~\ref{sec:conclusion} contains concluding remarks.

\section{Network growth via graph state-assisted measurements}
\label{sec:N-qubit_measurement}

\begin{figure}
\includegraphics[width=0.8\linewidth]{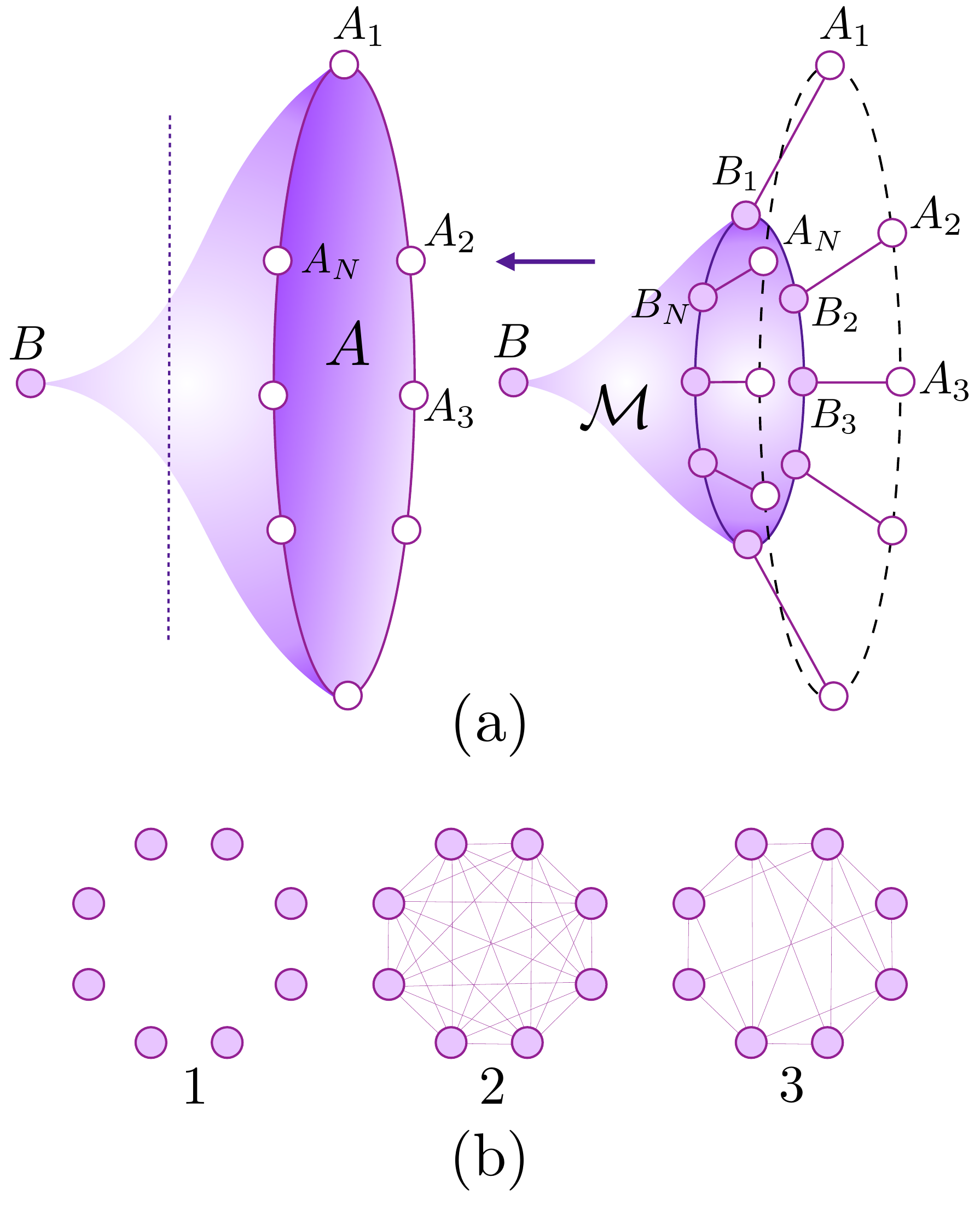}
\caption{(a) An $N$-qubit measurement $\mathcal{M}$  as defined in Eq.~(\ref{eq:measurement_operators}) on the $B$ qubits of $N$ non-maximally entangled qubit pairs $A_iB_i$ ($i=1,2,\cdots,N$)  projects the $N$-qubit Hilbert space $\otimes_{i=1}^N\mathbb{C}_{B_i}^2$ to a single-qubit Hilbert space $\mathbb{C}_B^2$. (b) Examples of (1) a fully disconnected, (2) a fully connected, and (3) an arbitrary graph of $8$ nodes are shown, with the corresponding graph states given by Eq.~(\ref{eq:graph_state}).  
}
\label{fig:schematic}
\end{figure}

In this section, we formally establish the scheme for network growth via graph state-assisted $N$-qubit measurements. Consider $N$ qubit-pairs $A_iB_i$ ($i=1,2,\cdots,N$), each referred to as \emph{unit} systems (see Fig.~\ref{fig:schematic}(a)), sharing an \emph{unit} entangled state
\begin{eqnarray}
\ket{\psi_i}=\cos\theta_i\ket{0_{B_i}}\ket{\tilde{0}_{A_i}}+\sin\theta_i\ket{1_{B_i}}\ket{\tilde{1}_{A_i}},\label{eq:ini_weak_states}
\end{eqnarray}
in $\mathbb{C}^2_{B_i}\otimes \mathbb{C}^2_{A_i}$, with $\{\ket{0},\ket{1}\}$ ($\{\ket{\tilde{0}},\ket{\tilde{1}}\}$) being the complete orthonormal basis in $\mathbb{C}^2_{B_i}$ ($\mathbb{C}^2_{A_i}$)~\cite{nielsen2010}, where we assume $\{\ket{0},\ket{1}\}$ to be the computational basis. The state $\ket{\psi_i}$ contains entanglement,
\begin{equation}
    E_i=\sin 2\theta_i,
\end{equation}
as quantified by concurrence~\cite{Hill1997} (c.f.~\cite{Coffman2000}), between $A_i$ and $B_i$, where $\theta_i\in[0,\pi/4]$ with $\theta_i=0$ representing an un-entangled state, and $\theta=\pi/4$ corresponds to a maximally entangled state. We refer to the set of $N$ qubits labeled by $A_i (B_i)$, $i=1,2,\cdots,N$, as the $A$-qubits ($B$-qubits). 

\subsection{Graph state-assisted measurement}
\label{subsec:graph_measurement}

We perform a joint measurement $\mathcal{M}$ on the $B$-qubits projecting the $N$-qubit Hilbert space $\otimes_{i=1}^N\mathbb{C}^{2}_{B_i}$ to a single-qubit one (see Fig.~\ref{fig:schematic}(a)), denoted by $\mathbb{C}^2_B$, leading to $(N+1)$-qubit post-measured states  in which the partition $A$ contains $N$ qubits. The measurement is  given by $2^{N-1}$ measurement operators 
\begin{eqnarray}    
M_{k}=\ket{0}\bra{+_{k}}+\ket{1}\bra{-_{k}},
\label{eq:measurement_operators}
\end{eqnarray}
corresponding to the measurement outcomes $k$,  where
\begin{equation}
\ket{\pm_{k}}=\frac{\ket{e_{b_{k}}}\pm\ket{e_{b_{k}^\prime}}}{\sqrt{2}},
\end{equation}
with $b_k\neq b^\prime_k$ $\forall k$. Here, $\set{\ket{e_{b_k}}}$ with  
\begin{eqnarray}
    \ket{e_{b_k}}=U_B^G\ket{b_k}=\bigg[ \prod_{(i,j)\in G} U^{\text{cz}}_{(i,j)} H_B \bigg] \ket{b_k}\label{eq:graph_state},
\end{eqnarray}
is the graph state basis~\cite{hein2004}, where $\{\ket{b_k}\}$ is the computational product basis on $B$-qubits. Note that for a fixed $k$, there are $2^N$ choices for $\ket{b_k}$, and therefore also for $\ket{e_{b_k}}$. In Eq.~(\ref{eq:graph_state}), $G$ is an arbitrary, undirected, simply connected graph on $N$ nodes without any loops connecting a node to itself~\cite{hein2006}, specified completely by the $N\times N$ adjacency matrix $\Gamma$ with elements,
\begin{eqnarray}
    \Gamma_{i,j}=\begin{cases}
        1, \forall i\neq j\text{ iff } (i,j)\in G,\\
        0, \text{ otherwise,}
    \end{cases}
\end{eqnarray}
where $(i,j)$ denotes a link between the nodes $i$ and $j$ (see Fig.~\ref{fig:schematic}(b)). The unitary operator, 
\begin{equation}
    U^{\text{cz}}_{(i,j)}=\frac{1}{2}\left[(I+\sigma^z)_{i}I_{j}+(I-\sigma^z)_{i}\sigma^z_{j}\right],
\end{equation}
is the controlled $Z$ operator acting on the qubit-pair $(i,j)$ $(i \neq j)$, and  $H_B=\otimes_{i=1}^NH_{B_i}$, $H_{B_i}$ being the Hadamard operators. 

In order to obtain the full set of $2^{N-1}$ measurement operators, we consider the following construction. First, note that any two distinct outcomes $k$ and $k^\prime$ can be connected by a unique bit-string $\ell$ of length $N-1$ as,
\begin{equation}
    k^\prime=k\oplus \ell,
\end{equation}
where unless otherwise specified, we always assume bit-wise modular addition between bit-strings. This leads to the connection 
\begin{eqnarray}\label{eq:all_pairs}
(b_{k^\prime},b_{k^\prime}^\prime)=(b_k\oplus \ell,b_k^\prime\oplus \ell)
\end{eqnarray}
between the bit-strings $(b_{k^\prime},b_{k^\prime}^\prime)$, defining the measurement operators (Eq.~(\ref{eq:measurement_operators})) corresponding to the outcomes $k$ and $k^\prime$ respectively, ensuring that there is no repetition of the pair of bit-strings, thereby resulting in the completeness $\sum_k M^\dagger_{k}M_{k}=I$. Therefore, the complete set $\{M_{k}\}$ can be generated by only varying $\ell\in\left[1,2^{N-1}-1\right]$ starting from $M_0$ corresponding to $k=0$. We refer to the pair of bit-strings $(b_0,b^\prime_0)$ corresponding to $M_0$ as the \emph{zero'th pair}. Note that $b_0\;(b^\prime_0)$ has $2^N\;(2^N-1)$ possibilities, and therefore we assume the choice of the zero'th pair to be arbitrary within this set. 

\subsection{Effectively deterministic generation of multi-qubit states}

We now discuss the growth of the network of qubits starting from $N$ qubit-pairs via the measurement discussed in Sec.~\ref{subsec:graph_measurement}. To establish that the growth of the network is deterministic despite the use of measurement, and to understand the entanglement property of the network generated, the following proposition is crucial. 

\begin{figure}
    \centering
    \includegraphics[width=0.8\linewidth]{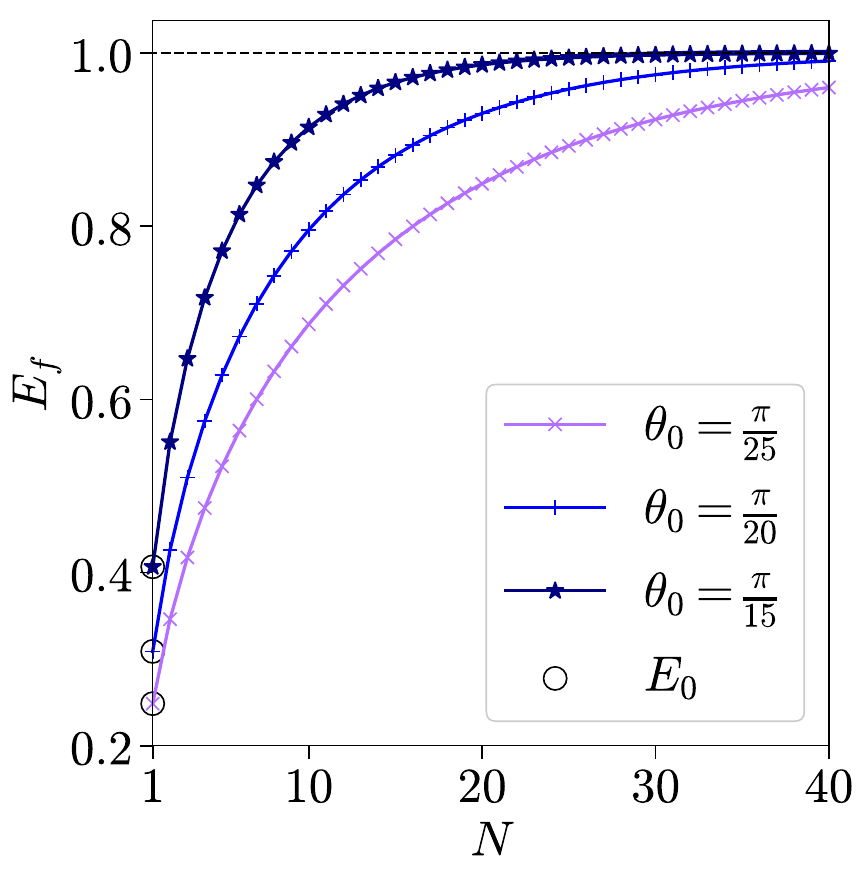}
    \caption{Variations of $E_f$ (Eq.~(\ref{eq:Ef_arbitrary_N_formula})) as a function of $N$ for different $E_0$ (i.e., different $\theta_0$), where $E_i=E_0\forall i$.}
    \label{fig:main_results}
\end{figure}

\proposition{1} \emph{The $N$-qubit measurement $\mathcal{M}$ (Eq.~(\ref{eq:measurement_operators})) leads to equiprobable measurement outcomes $k$ and local unitary-equivalent post-measured states $\ket{\Phi_{k}}$, where the unitary operators are of the form $U_A\otimes U_B$, thereby making the measurement effectively deterministic as far as bipartite entanglement over the partition $A:B$ is concerned}.

\begin{figure*}
    \centering
    \includegraphics[width=0.8\linewidth]{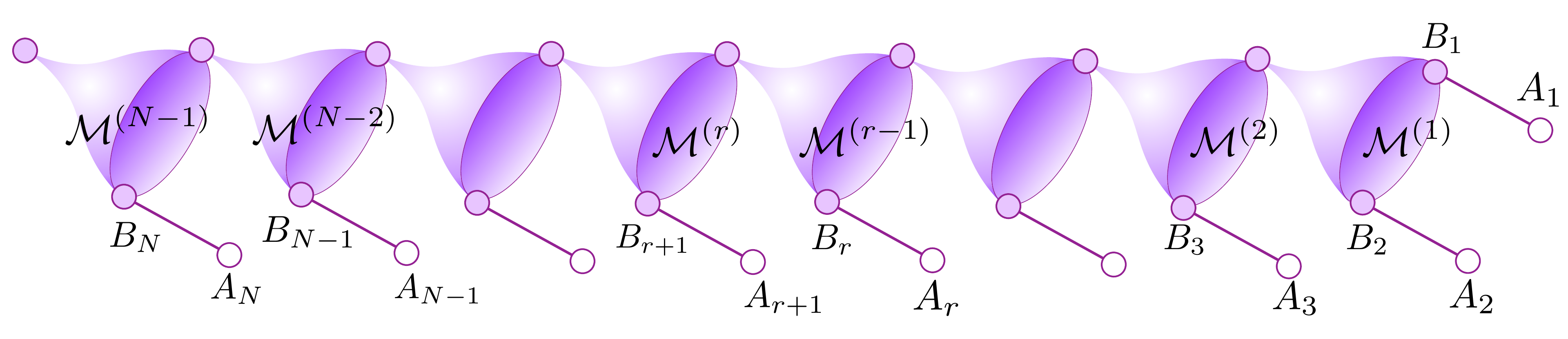}
    \caption{Repetitive two-qubit measurement scheme with $N-1$ rounds of measurements applied to the $N$ qubit-pairs in the order $1,2,\cdots,N$.}
    \label{fig:two_qubit_scheme}
\end{figure*}

\begin{proof} The bipartite system $A:B$ is in the joint state  
\begin{eqnarray}
    \ket{\Psi}=\sum_{b=0}^{2^N-1}c_b \ket{b}\ket{\chi^{b}_{A}},
    \label{eq:composite_computational_supple}
\end{eqnarray}
with
\begin{equation}\label{eq:sum_c_b_sq}
    \sum_{b=0}^{2^N-1} c_{b}^2=1,
\end{equation}
where
\begin{equation}
    \ket{b}\equiv\otimes_{i=1}^N\ket{b_{B_{i}}}
\end{equation}
being the computational basis states such that $b_{B_i}=0_{B_i},1_{B_i}$. Also,
\begin{equation}
\ket{\chi_{A}^{b}}=\otimes_{i=1}^N\ket{\widetilde{(b_{B_i})}_{A_i}},
\end{equation}
where $\widetilde{(b_{B_i})}_{A_i}=\tilde{0}_{A_i},\tilde{1}_{A_i}$  and,
\begin{equation}
    c_b=\prod_{i=1}^N (\cos\theta_{i})^{1\oplus b_{B_i}}(\sin\theta_i)^{b_{B_i}}.
\end{equation}
Here $\oplus$ denotes  the modular two addition. Now consider the joint measurement $\mathcal{M}$ given by the measurement operators Eq.~(\ref{eq:measurement_operators}) involving the graph basis states. For an arbitrary, un-directed, simply connected graph $G$ with adjacency matrix $\Gamma$, the following matrix elements can be written,
\begin{equation}\label{eq:matrix_element_arb_graph}
\bracket{e_b}{b^\prime}=\langle b|U_B^G|b^\prime\rangle=2^{-N/2}(-1)^{f_{(b,b^\prime)}+\gamma_b},   
\end{equation}
where the binary values,
\begin{eqnarray}
    f_{(b,b^\prime)}&=&\bigoplus_i b_{B_i}\wedge b_{B_i}^\prime \in\{0,1\},\label{eq:f_b_bprime}\\
    \gamma_{b}&=&\bigoplus_{i,j=1}^N \Gamma_{i,j} b_{B_i} b_{B_j} \in\{0,1\}
\end{eqnarray}
with $\wedge$ denoting the logical AND operation between two bits $b_{B_i}$ and $b_{B_i}^\prime$. The action of $M_{k}$ on $\ket{\Psi}$ (Eq.~(\ref{eq:composite_computational_supple})) results in 
\begin{eqnarray}\label{eq:post_measured_state_intermediate}
    M_{k} \ket{\Psi}&=&2^{-\frac{N-1}{2}}\bigg(\sum_b c_b \big[ g^+_b\ket{0_B} +g^-_b\ket{1_B}\big] \ket{\chi^{b}_A}\bigg),\nonumber\\ 
\end{eqnarray}
with 
\begin{eqnarray}\label{eq:gb_pm}
g^{\pm}_b=\frac{(-1)^{f_{(b_k,b)}+\gamma_{b_k}}\pm(-1)^{f_{(b_k^\prime,b)}+\gamma_{b^\prime_k}}}{2},
\end{eqnarray}
such that $g^{\pm}_b=\pm 1$ implies $g^{\mp}_b=0$. Therefore, when combined with Eq.~(\ref{eq:sum_c_b_sq}), it is easy to see that the state within the parenthesis of Eq.~(\ref{eq:post_measured_state_intermediate}) is already normalized yielding the probability 
\begin{eqnarray}
    p_k=\langle\Psi|M_{k}^\dagger M_{k}|\Psi\rangle=2^{-(N-1)}\label{eq:equiprobable},
\end{eqnarray}
to obtain an outcome $k$, implying that all outcomes are equiprobable. 

Next, note that $k^\prime=k\oplus \ell$ is equivalent to local sign flips of $\ket{e_b}$, i.e., $\ket{e_{b\oplus \ell}}=Z^{(\ell)}_B\ket{e_b}$, where $Z_B^{(\ell)}=\prod_{i=1}^N\left(\sigma^z_{B_i}\right)^{\ell_{B_i}}$, $\ell_{B_i}\in\{0,1\}$ being the bit corresponding to $B_i$ in $\ell$. Note also that 
\begin{eqnarray}
    \bracket{e_{b\oplus \ell}}{b^\prime}=2^{-N/2}(-1)^{f_{(b,b^\prime)}+\gamma_b}(-1)^{f_{(\ell,b^\prime)}},
\end{eqnarray}
which leads to the normalized post measured state corresponding to another outcome $k^\prime$,
\begin{eqnarray}\label{eq:post_measured_state_kprime} 
    \ket{\Phi_{k^\prime}}&=&\sum_b (-1)^{f_{(\ell,b)}} c_b \big[ g^+_b\ket{0_B} +g^-_b\ket{1_B}\big] \ket{\chi^{b}_A},
\end{eqnarray}
implying introduction of a local phase with the $b$th term when $f_{(\ell,b)}=1$. A unitary operator of the form $U^{\ell}_{AB}=I_B\otimes U^{\ell}_A$, which is local to the partitions $A$ and $B$ with $I_B$ being the identity matrix on $B$, and
\begin{equation}\label{eq:phase_negate_unitary}
U^{\ell}_{A}= \sum_b(-1)^{f_{(\ell,b)}}\ket{\chi^{b}_A}\bra{\chi^{b}_A},    
\end{equation}
negates these local phases, leading to 
\begin{eqnarray}\label{eq:lu_equivalence}
\ket{\Phi_{k^\prime}}=U^{\ell}_{AB}\ket{\Phi_{k}}.
\end{eqnarray}
This, along with Eq.~(\ref{eq:equiprobable}), proves Proposition 1. 
\end{proof} 

\noindent\textbf{Note 1.} Note that the Proposition 1 remains valid for arbitrary graphs due to the fact that a change in connectivity of the graph only results in a change in the value of $\gamma_b$, therefore affecting neither Eq.~(\ref{eq:equiprobable}) nor Eq.~(\ref{eq:lu_equivalence}). Without any loss in generality, in what follows, we will consider a completely disconnected graph (see Fig.~\ref{fig:schematic}(b)), for which $U_B^G=I$, and $\gamma_b=0\forall b$.

\subsection{Bipartite entanglement in post-measured states} 

The effectively deterministic nature of $\mathcal{M}$ with respect to the bipartite entanglement over the partition $A:B$ in the post-measured states allows one to consider only $\ket{\Phi_{0}}$, corresponding to $k=0$, in all subsequent discussions. Using Eq.~(\ref{eq:post_measured_state_kprime}), $\ket{\Phi_{0}}$ takes the Schmidt decomposed form
\begin{eqnarray}\label{eq:final_Schmidt_decomposed}
\ket{\Phi_{0}}=\cos\theta_f\ket{0_{B}}\ket{\mathbf{0}_A}+\sin\theta_f\ket{1_{B}}\ket{\mathbf{1}_A}.
\end{eqnarray}
where
\begin{equation}\label{eq:cos_thetaf_supple}
    \cos\theta_f=\sqrt{\sum_{b\in \mathcal{B}}c_b^2},
\end{equation}
and $\{\ket{\mathbf{0}_A},\ket{\mathbf{1}_A}\}$ are multi-qubit states on the $A$-qubits having the form 
\begin{equation}
    \ket{\mathbf{0}_A}=\sum_{b\in \mathcal{B}}c_b\ket{\chi_A^b},\hspace{1cm}
    \ket{\mathbf{1}_A}=\sum_{b\in \overline{\mathcal{B}}}c_b\ket{\chi_A^b},
\end{equation}
with $\langle\mathbf{0}|\mathbf{1}\rangle=0$. Here we have defined the mutually exclusive sets of bit-strings
\begin{equation}\label{eq:set_B_Bbar_supple}
    \mathcal{B}=\{b:g^+_b\neq0\},\hspace{1cm}\overline{\mathcal{B}}=\{b:g^-_b\neq0\}.
\end{equation}
Evidently,  $\theta_f$ depends explicitly on the choice of the zero'th pair $(b_0,b_0^\prime)$ via Eqs.~(\ref{eq:gb_pm}) and (\ref{eq:f_b_bprime}). For a specific choice of the zero'th pair, the  bipartite entanglement over the partition $A:B$ is given by $E_f=\sin 2\theta_f$ for all $k$. 

\subsubsection{Profitable concentration of bipartite entanglement}
\label{subsubsec:PCE}

In general, the bipartite entanglements of individual pairs participating in the measurement can  be different, i.e., $E_i\neq E_j$ for $i\neq j$, and it is desirable to achieve
\begin{eqnarray}
    E_f\geq E_0= \max\{E_i;i=1,2,\cdots,N\}
    \label{eq:pce}
\end{eqnarray}
for at least one choice of the zero'th pair --  a situation corresponding to a  \emph{profitable concentration of entanglement} (PCE), when bipartite entanglement over $A:B$ is considered as resource. Here, we consider the case of all the pairs participating in the measurement having identical entanglements $E_i=E_0\forall i$, where the following proposition establishes PCE.

\proposition{2} \emph{For $N$ identical qubit-pairs, each having bipartite entanglement $E_0$ $(0\leq E_0\leq 1)$, 
\begin{eqnarray}
    E_f=\sqrt{1-(1-E_0^2)^N}\geq E_0
    \label{eq:entanglement_simplified}
\end{eqnarray}
for all $E_0$ with the choice of the zero'th pair
\begin{eqnarray}\label{eq:zeroth_pair_main}
(b_0,b_0^\prime)=(00\cdots0,11\cdots1).
\end{eqnarray}}

\begin{proof} 

The choice of zeroth pair $(b_0,b_0^\prime)=(00\cdots 0,11\cdots 1)$ leads to the following form of the sets $\mathcal{B}$ and $\overline{\mathcal{B}}$:
\begin{eqnarray}
\mathcal{B}&=&\{b;\oplus_i b_i=0 \},\; 
\overline{\mathcal{B}}=\{b;\oplus_i b_i=1 \}. \label{eq:set_B}
\end{eqnarray}
Note that in this case, $\mathcal{B}$ ($\overline{\mathcal{B}}$) is the \emph{even-parity set} (\emph{odd-parity set}), where the parity is the number $1$'s in the corresponding bit-string $b$ for each element. Also not that the cardinality of the sets are $|\mathcal{B}|=|\overline{\mathcal{B}}|=2^{N-1}$. Assuming identical bipartite systems $A_iB_i$ with $\theta_i=\theta_0\forall i$ in Eq.~(\ref{eq:cos_thetaf_supple}), one can obtain $\cos\theta_f=\sqrt{\left[1+(\cos 2\theta_0)^N\right]/2}$, leading to
\begin{eqnarray}
    E_f=\sqrt{1-(1-E_0^2)^N}
    \label{eq:Ef_arbitrary_N_formula},
\end{eqnarray}
where $E_0=\sin2\theta_0$. Note that $E_f\geq E_0$ for all $0\leq E_0\leq 1$. Hence the proof.
\end{proof} 

In Fig.~\ref{fig:main_results} we demonstrate the variation of $E_f$ with $N$ for different values of $E_0$ showing the monotonic increase in $E_f$ with number of entangled pairs $N$ participating in the measurement. Note that for a fixed $E_0>0$, $E_f$ increases monotonically with $N$: $E_f\rightarrow 1$ as $N\rightarrow\infty$ $\forall$ $E_0>0$, or as $E_0\rightarrow 1$ $\forall$ $N$. Further, the number $m_N$ of qubit-pairs with  entanglement $E_0$ required to obtain $E_f$, as given in Eq.~(\ref{eq:entanglement_simplified}), is
\begin{equation}
   m_N=\bigg\lceil\frac{\ln(1-E_f^2)}{\ln(1-E_0^2)}\bigg\rceil,\label{eq:mJ}
\end{equation} 
where $\lceil .\rceil$ is the ceiling of a real number. For a discussion on achieving the PCE via a scalable two-qubit repetitive measurement scheme, see Sec.~\ref{subsec:achieving_PCE}.

\section{Scalable growth via repeated two-qubit measurements}
\label{sec:repeated_two_qubit_measurement}

Note that the joint N-qubit measurement proposed in Eq.~(\ref{eq:measurement_operators}) lacks scalability as it demands a case by case construction of associated measurement procedure for a given $N$. As an alternative but scalable approach, in this section, we propose a repeated two-qubit measurement scheme (see Fig.~\ref{fig:two_qubit_scheme}). 

\begin{figure}
    \centering
    \includegraphics[width=0.8\linewidth]{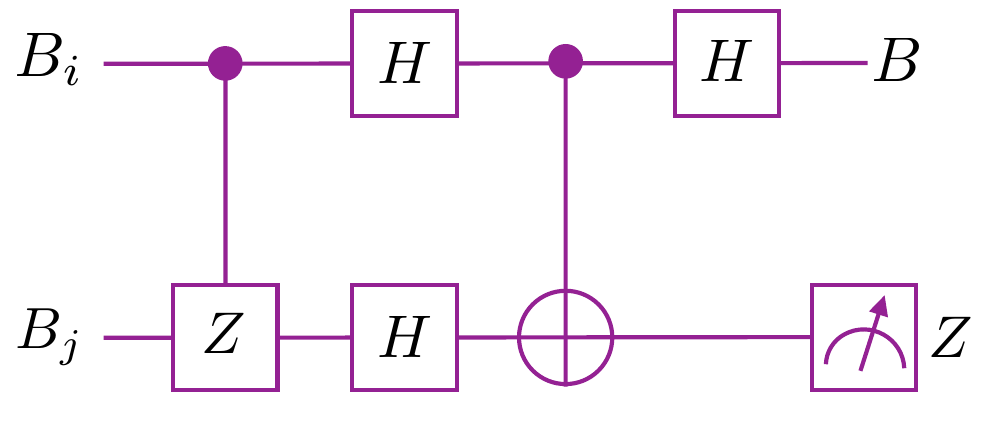}
    \caption{Execution of the two-qubit measurement (Eq.~(\ref{eq:star_twoqubit_meas})).}
    \label{fig:two_qubit_measurement_circuit}
\end{figure}

Let us first specifically focus on the case of $N=2$, where the two-qubit measurement $\mathcal{M}$ on the $B$-qubits is defined by the measurement operators given in Eq.~(\ref{eq:measurement_operators}). With the choice of zeroth pair $(b_0,b_0^\prime)=(00,11)$ (see Eq.~(\ref{eq:zeroth_pair_main}); see Appendix~\ref{app:two_qubit_measurement} also for other choices of zeroth pairs), the measurement operators corresponding to outcomes $k=0,1$ are given by
\begin{eqnarray}\label{eq:star_twoqubit_meas}    M_{0}&=&\ket{0}\left(\frac{\bra{e_{00}}+\bra{e_{11}}}{\sqrt{2}}\right)+\ket{1}\left(\frac{\bra{e_{00}}-\bra{e_{11}}}{\sqrt{2}}\right),\nonumber\\
M_{1}&=&\ket{0}\left(\frac{\bra{e_{01}}+\bra{e_{10}}}{\sqrt{2}}\right)+\ket{1}\left(\frac{\bra{e_{01}}-\bra{e_{10}}}{\sqrt{2}}\right).
\end{eqnarray}
This, on any two $B$-qubits, say, $B_i$ and $B_j$, can be performed via a two-qubit parity measurement preceded by an appropriate basis transformation given by  $(\ket{e_{00}}+\ket{e_{11}})/\sqrt{2}\to \ket{00}$, $(\ket{e_{00}}-\ket{e_{11}})/\sqrt{2}\to \ket{11}$, $(\ket{e_{01}}+\ket{e_{10}})/\sqrt{2}\to \ket{01}$, and $(\ket{e_{01}}-\ket{e_{10}})/\sqrt{2}\to \ket{10}$,  brought about by the two-qubit controlled $Z$ operation and single-qubit Hadamard gates (see Fig.~\ref{fig:two_qubit_measurement_circuit}). After the controlled NOT operation, the target qubit is measured in the computational basis, resulting in either $+1$, corresponding to $M_0$, or $-1$, corresponding to $M_1$. Post parity measurement, the control qubit $B_i$ remains entangled with the parties $A_i$ and $A_j$, and can be re-labelled as $B$ post a Hadamard operation,  while the target qubit $B_j$ is fully disentangled, and therefore, can be discarded.

\subsection{Two-qubit repeated measurement scheme}
\label{subsec:two_qubit_scheme}

The repeated two-qubit  measurement scheme proceeds as follows. In the first step, the two-qubit measurement $\mathcal{M}^{(1)}$ 
is applied on the $B$ qubits of the entangled qubit-pairs  in states $\ket{\psi_1}$ and $\ket{\psi_2}$ to create entangled post-measured states $\{\ket{\Phi_{0}^{(1)}}, \ket{\Phi_{1}^{(1)}}\}$, that are connected via local unitary operations (Proposition 1). Next, the second round of two-qubit measurement, $\mathcal{M}^{(2)}$, can be performed on any one of the states $\ket{\Phi_{k}^{(1)}}$, $k=0,1$, and the state $\ket{\psi_3}$. This process can be continued (see Fig.~\ref{fig:schematic}(c)) $r$ times, taking $(r+1)$ two-qubit states $\ket{\psi_i}$, $i=1,2,3,\cdots,r+1$, resulting in a tree of $2^r$ paths leading to $2^r$ possible post-measured states after the completion of $r$ rounds of measurements. The final ensemble post $\mathcal{M}^{(r)}$ is decided by the selection of $k$ post each round of measurements. 

\begin{figure*}
    \centering
    \includegraphics[width=0.8\linewidth]{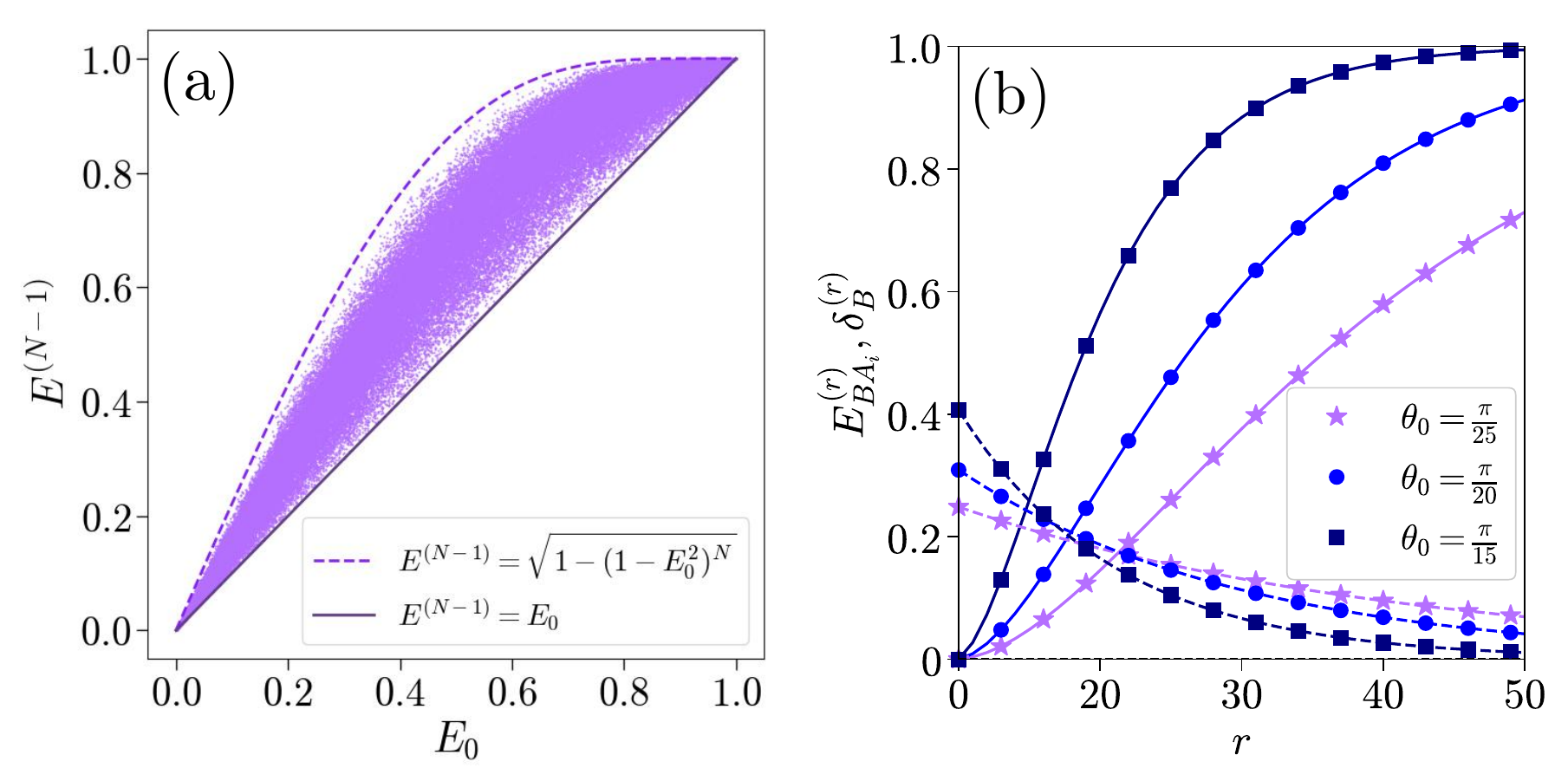}
    \caption{ (a) Scatter plot of $E^{(N-1)}$ and $E_0$ with $N=5$ demonstrating the bounds on $E^{(N-1)}$ given by Eq.~(\ref{eq:bounds}), where $E_0=\max\{E_i\}$, and $E_i$s are generally different.  A sample of size $10^5$ instances with different states $\ket{\psi_i}$ having different $E_i$ is used, where $E_0$ for each instance is sampled from a uniform distribution over $[0,1]$. (b) Variations of $E_{B:A_i}^{(r)}$ (dashed lines, Eq.~(\ref{eq:EBAi})) as quantified by concurrence, and $\delta_B^{(r)}$ (continuous lines, Eq.~(\ref{eq:MS})) quantified by the squared concurrence, as functions of $r$.
    }
    \label{fig:bounds}
\end{figure*}

Outcome of this scheme is connected to the outcome of the $N$-qubit measurement (Eq.~(\ref{eq:measurement_operators})) discussed in Sec.~\ref{sec:N-qubit_measurement} via the following proposition.

\proposition{3} \emph{Using $k=0$ post each round of two-qubit measurements for the next round of measurement
results in $\ket{\Phi_{0}^{(N-1)}}=\ket{\Phi_{0}}$ (Eq.~(\ref{eq:final_Schmidt_decomposed})).}\footnote{Note here that due to Proposition 1, one can work with any value of $k$. While the details of the states will change, entanglement properties will remain unchanged.}

\begin{proof} 
We closely look at the $r$-th round, $\mathcal{M}^{(r)}$, of the repetitive two-qubit measurement corresponding to the chosen zero'th pair. While performed on the qubits belonging to subsystem $B$ of (a) the state $\ket{\Phi_0^{(r-1)}}$ in $\mathbb{C}_{B}^{2}\otimes \mathbb{C}_{A_{(r-1)}}^{2^r}$ post the $(r-1)$th rounds of two-qubit measurement corresponding to $M_0$, with $A_{(r-1)}=\bigcup_{i=1}^{r}A_i$ and (b) the state $\ket{\psi_{r+1}}$ in $\mathbb{C}_{B_{r+1}}^2\otimes \mathbb{C}_{A_{r+1}}^2$, such that the ensemble   $\left\{\ket{\Phi_0^{(r)}}, \ket{\Phi_1^{(r)}}\right\}$ in $\mathbb{C}_{B}^2\otimes \mathbb{C}_{A_{(r)}}^{2^{r+1}}$, with $A_{(r)}=A_{(r-1)}\cup A_{r+1}$ is obtained. One can write the post measurement state corresponding to the outcome zero: 
\begin{eqnarray}\label{eq:post_measured_state_r_rounds}
    \ket{\Phi_0^{(r)}}&=&\ket{0_B} \sum_{b\in\mathcal{B}_{(r)}} c_b \ket{\chi^b_{A_{(r)}}}+\ket{1_B} \sum_{b\in\overline{\mathcal{B}}_{(r)}} c_b \ket{\chi^b_{A_{(r)}}},\nonumber\\
\end{eqnarray}
where the sets $\mathcal{B}_{(r)}$ and $\overline{\mathcal{B}}_{(r)}$, defined in the same vein as in Eq.~(\ref{eq:set_B_Bbar_supple}) but this time corresponding to the $r-$th round, and is also given by the recurrence relation,  
\begin{eqnarray}
    \mathcal{B}_{(r)}&=&\left(\mathcal{B}_{(r-1)}\circ \mathcal{B}_{(0)}\right) \cup \left(\overline{\mathcal{B}}_{(r-1)}\circ \overline{\mathcal{B}}_{(0)}\right),\\
    \overline{\mathcal{B}}_{(r)}&=&\left(\mathcal{B}_{(r-1)}\circ \overline{\mathcal{B}}_{(0)}\right) \cup \left(\overline{\mathcal{B}}_{(r-1)}\circ \mathcal{B}_{(0)}\right).\label{eq:Br_evolution}
\end{eqnarray}
Here we define the zero'th round sets as $\mathcal{B}_{(0)}=\{0\},\overline{\mathcal{B}}_{(0)}=\{1\}$ and the operation $\mathcal{L}\circ\mathcal{R}$ between the sets $\mathcal{L}=\{L_1,L_2,\cdots L_n\}$ and $\mathcal{R}=\{R_1,R_2,\cdots R_n\}$ of bit-strings yields a new set of bit-strings by concatenating each element in $\mathcal{L}$ with that of $\mathcal{R}$ i.e, $\mathcal{L}\circ\mathcal{R}=\{L_1R_1,L_1R_2,\cdots L_nR_{n-1},L_nR_n\}$. For an example, 
\begin{eqnarray}
    \overline{\mathcal{B}}_{(1)}&=&\left(\{0\}\circ \{1\} \right) \cup \left(\{1\}\circ \{0\}\right)\nonumber\\
    &=&\{01,10\},\nonumber\\
    \overline{\mathcal{B}}_{(2)}&=&\left(\{00,11\}\circ \{1\} \right) \cup \left(\{01,10\}\circ \{0\}\right)\nonumber\\
    &=&\{001,010,100,111\},    
\end{eqnarray}
and so on. This process can be iterated $r$ times to obtain $\mathcal{B}_{(r)}$ as well as $\overline{\mathcal{B}}_{(r)}$.
The cardinality of the sets being  $|\mathcal{B}_{(r)}|=|\overline{\mathcal{B}}_{(r)}|=2^{r}$. Note that the recursion relation in Eq.~(\ref{eq:Br_evolution}) preserves the parity of $\mathcal{B}_{(r)}$ and $\overline{\mathcal{B}}_{(r)}$ to be even and odd, respectively. One can identify them to be the sets given in Eq.~(\ref{eq:set_B}) i.e, $\mathcal{B}_{(r)}=\mathcal{B}$ and $\overline{\mathcal{B}}_{(r)}=\overline{\mathcal{B}}$, when $r=N-1$, which corresponds to the joint $N$-qubit measurement. Thus, $\ket{\Phi_0^{(N-1)}}$, as given in Eq.~(\ref{eq:post_measured_state_r_rounds}) after $r=N-1$ rounds of two-qubit measurement, is equal to $\ket{\Phi_0}$ in Eq.~(\ref{eq:final_Schmidt_decomposed}). Hence the proof. 
\end{proof}

\noindent\textbf{Note 2.} It is important to mention here that the $2^r$ possible post-measured states can be identified as the $2^{N-1}$ post-measured states corresponding to the $N$-qubit joint measurement when $r=N-1$, while it is easy to see from Proposition 1 that  each of these $2^r$ states are connected to each other by local unitary operations of the form $U_A\otimes U_B$, thereby having the same bipartite entanglement over the partition $A:B$. 

\subsection{Achieving profitable concentration of entanglement}
\label{subsec:achieving_PCE}

Let us denote the bipartite entanglement between the $B$ qubit and the rest of the qubits in $\ket{\Phi_{0}^{(r-1)}}$ by $E^{(r)}$, and the entanglement between the qubits $A_{r+1}$ and $B_{r+1}$ in the qubit-pair $A_{r+1}B_{r+1}$ in $\ket{\psi_{r+1}}$ by  $E_{r+1}$. The following proposition is to show that each round of measurements on the available $N$ qubit-pairs always leads to PCE. 

\proposition{4} \emph{For two consecutive rounds $r+1$ and $r$ of two-qubit measurements, $E^{(r+1)}\geq E^{(r)}$ $\forall r$. }

\begin{proof} 
Combining the pairs with entanglement $E^{(r)}$ and
$E_{r+1}$ via two-qubit measurement, the $A:B$ bipartite entanglement of the post measured state $\ket{\Phi_{0}^{(r)}}$ after the $r$'th round of two-qubit measurement $\mathcal{M}^{(r)}$ can be obtained to be (see Appendix~\ref{app:two_qubit_measurement}):
\begin{eqnarray}
        {E^{(r+1)}}&=&\sqrt{(E^{(r)})^2 + E_{r+1}^2-(E^{(r)})^2 E_{r+1}^2}.
        \label{eq:Er+1}
\end{eqnarray}
Without any loss of generality, we assume $E_{r+1}=\max\{E^{(r)},E_{r+1}\}$ and $E^{(r)}=q_rE_{r+1}$ with $0\leq q_r\leq 1$. Thus, 
\begin{eqnarray}
        {E^{(r+1)}}&=&E_{r+1}\sqrt{1+(1-E_{r+1}^2)q_r^2}.
        \label{eq:Er+1_modified}
\end{eqnarray}
Since $\sqrt{1+(1-E_{r+1}^2)q_r^2}\geq 1$, $E_{(r+1)}\geq E_{r+1}\geq E_{(r)}$. Similarly one can again show, $E^{(r+1)}\geq E^{(r)}$ when $E^{(r)}=\max\{E^{(r)},E_{r+1}\}$ as well. Hence the proof. 
\end{proof}

Let us now consider the situation of $N$ qubit-pairs in states $\ket{\psi_i}$ with bipartite entanglements as $E_i$, $i=1,2,\cdots,N$, where one performs $N-1$ rounds of repetitive two-qubit measurements merging the qubit-pairs one-by-one in the order $1,2,\cdots,N$, and obtains the bipartite entanglement in the final $(N+1)$-qubit system over the bipartition $A:B$ as $E^{(N-1)}$. The next proposition provides bounds on $E^{(N-1)}$ (see Fig.~\ref{fig:bounds}(a) for a demonstration).  

\proposition{5} \emph{The bipartite entanglement $E^{(N-1)}$ satisfies 
\begin{eqnarray}
    E_0\leq E^{(N-1)} \leq \sqrt{1-\left(1-E_0^2\right)^N},
    \label{eq:bounds}
\end{eqnarray}
where $E_{0}=\max\{E_1,E_2,\cdots,E_N\}$.}

\begin{proof}
To prove this, given that $E_{0}=\max\{E_1,E_2,\cdots,E_N\}$, without any loss in generality, we assume $E_i=q_iE_0\forall i$ with $0\leq q_i\leq 1$, and write $E^{(r+1)}$ as 
\begin{eqnarray}
    E^{(r+1)}=\sqrt{(E^{(r)})^2+q_{r+1}^2E_0^2\left(1-(E^{(r)})^2\right)},
\end{eqnarray}
which is a monotonically increasing function of $q_{r+1}$ for a fixed $E_0$ and $E^{(r)}$, attaining maximum at $q_{r+1}=1$. This also applies to all rounds of measurements, such that $q_{i}=1\forall i$ for $E^{(N-1)}$ to attain its maximum, leading to $E_i=E_0$ $\forall i$. Further, using proposition 2, the concentrated entanglement in this situation is given by Eq.~(\ref{eq:Ef_arbitrary_N_formula}), thereby proving the upper bound. Also, $(N-1)$ rounds of two-qubit measurements includes all $N$ two-qubit pairs, which, along with proposition 4 leads to the lower bound. Hence the proof.
\end{proof}

\noindent\textbf{Note 3.} It is also worthwhile to point out here that due to Schmidt decomposition, all of the above results remain valid if one replaces the $A$-qubits with qudits. This, in turn, allows multi-qubit states, such as the GHZ and the W states, as potential inputs to our protocol. While the details of the output states have to be worked out,  most of the results presented in this paper would qualitatively remain valid.  

\noindent\textbf{Note 4.} Evidently, the application of the repetitive two-qubit measurement scheme on identical two-qubit states each having entanglement $E_0$ results in monotonic increase of the bipartite entanglement $E_{B:A}=E^{(r)}$. On the other hand, the bipartite entanglement between $B$ qubit and one of the $A$ qubit (say $A_i$), denoted as $E^{(r)}_{BA_i}$, can be found to decrease and eventually disappear as $r\to\infty$ (see Appendix~\ref{app:two_qubit_unit_states} for details) as
\begin{eqnarray}\label{eq:EBAi}
    E_{BA_i}^{(r)}=E_0(1-E_0^2)^{r/2}.
\end{eqnarray}
Also, since $A_i, A_j$ belong to two disjoint states before measurement, and the two-qubit measurements are local to the subsystem $B$, one can easily see that the bipartite entanglement between the respective qubits, $E^{(r)}_{A_iA_j}=0,\forall A_i,A_j\in A$ where $i\neq j$. Using these results, one can see that the resulting multi-qubit state becomes increasingly monogamous, as quantified by the monogamy score~\cite{Dhar2017,*Zong2022}, $\delta_B^{(r)}$ (see Appendix.~\ref{app:two_qubit_unit_states} for definition and details), given by 
\begin{eqnarray}\label{eq:MS}
    \delta_B^{(r)}&=&1-(1-E_0^2)^{r}(1+rE_0^2),
\end{eqnarray}
quantified using the squared concurrence~\cite{Hill1997,Wootters1998,Wootters2001} (cf.~\cite{Coffman2000}), as $r$ increases, when the qubit $B$ is chosen as the nodal observer. In Fig.~\ref{fig:bounds}(b), we demonstrate this by plotting $E_{BA_i}^{(r)}$ and  $\delta_B^{(r)}$. Further, the state has a constant genuine multiparty entanglement (see Appendix~\ref{app:two_qubit_unit_states}), as quantified by the generalized geometric measure~\cite{sende2010a,*sende2012} (cf.~\cite{SHIMONY1995,*Barnum_2001,*Wei2003}).  

\section{Effect of imperfections on the protocol}
\label{sec:imperfections}

In this Section, we investigate the effect of possible imperfections on the protocol. For this, we focus on two specific cases, namely, the presence of noise in the two-qubit measurement, and the initial two-qubit states subjected to noise.

\begin{figure*}
    \centering
    \includegraphics[width=0.8\linewidth]{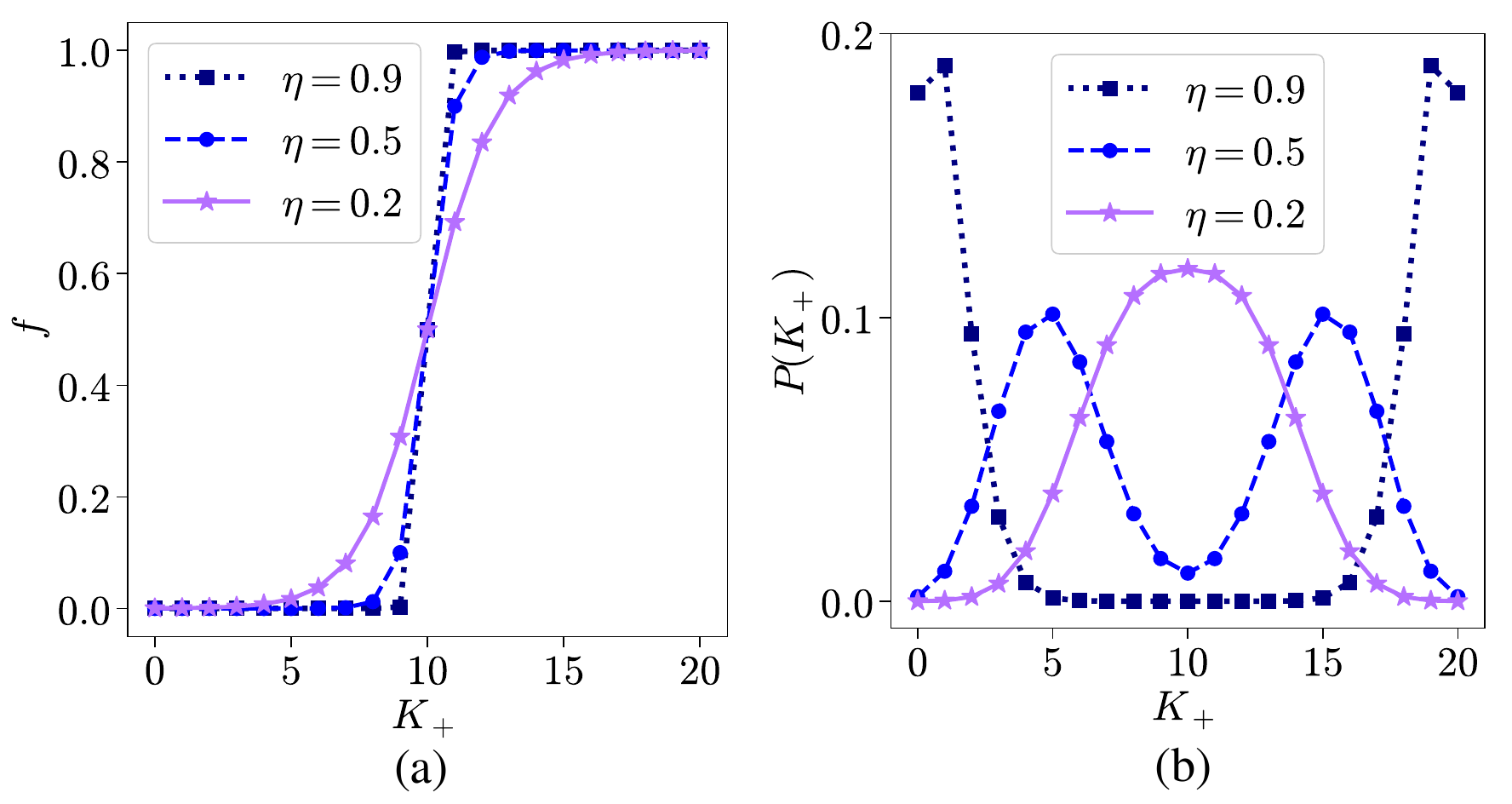}
    \caption{Plots of (a) $f$ (Eq.~\ref{eq:f}) and (b) $P(K_+)$ (Eq.~\ref{eq:P_K+}) against $K_+$ are shown for different values of $\eta$, where in both cases, $K$ is fixed at $20$ for demonstration.}
    \label{fig:noisy_measurement}
\end{figure*}

\subsection{Noisy two-qubit measurements}
\label{subsec:noisy_measurements}

The measurement formulation discussed so far assumes perfect execution of the measurement $\mathcal{M}$ via the circuit in Fig.~\ref{fig:two_qubit_measurement_circuit}, while in reality, it may be noisy due to a number of imperfections during experimental realizations, such as noisy gates and single-qubit measurement. Here we consider the  noise in the measurement apparatus in Fig.~\ref{fig:two_qubit_measurement_circuit} that manifest through the POVM elements~\cite{nielsen2010,Oreshkov2005}
\begin{eqnarray} \label{eq:unsharp_meas}   \mathcal{P}_\pm&=&\sqrt{(1\pm\eta)/2}\ket{0}\bra{0}+\sqrt{(1\mp\eta)/2}\ket{1}\bra{1},
\end{eqnarray}
corresponding to the outcomes $\pm 1$. Here, $\eta$ ($0\leq \eta\leq 1$) is the noise parameter with $\eta=0$ ($\eta=1$) corresponding to a completely noisy (noiseless) measurement. In such a scenario, to increase the fidelity of the post-measured state, we propose extending the protocol by repeatedly measuring the $B_j$ qubit $K$ times (c.f.~\cite{Mondal2025}), out of which the outcome $\pm 1$ is obtained $K_\pm$ times ($K_++K_-=K$), with the probability of obtaining $K_+$ of the outcomes as $+1$ given by 
\begin{eqnarray}\label{eq:P_K+}
    P(K_+)=\frac{1}{2}\genfrac(){0pt}{1}{K}{K_+}(1+\gamma)^{-K}\left[\gamma^{K_+}+\gamma^{K-K_+}\right],
\end{eqnarray}
where,
\begin{equation}
    \gamma=\frac{1-\eta}{1+\eta}.
\end{equation}
Post the $K$th measurement, the normalized post-measured state on $BA$ for $\eta<1$ is given by,
\begin{eqnarray}    \rho^K_{BA}=f\ket{\Phi_{0}}\bra{\Phi_{0}}+(1-f)\ket{\Phi_{1}}\bra{\Phi_{1}},
\end{eqnarray}
where, 
\begin{eqnarray}\label{eq:f}
    f&=&\big[1+\gamma^{(2K_+-K)}\big]^{-1},
\end{eqnarray}
is the fidelity of obtaining the target state $\ket{\Phi_0}$.  Note that $K_+\to K$ ($K_+\to 0$) corresponds to $f\to 1$ ($f\to 0$) irrespective of the value of $\eta$, while for $K_+=K/2$, $f=1/2$, leading to an equal mixture of  $\ket{\Phi_0}$ and $\ket{\Phi_1}$. In Figs.~\ref{fig:noisy_measurement}(a) and  (b), we respectively depict $f$ and $P(K_+)$ against $K_+$, which  clearly indicates that even with a relatively low $\eta$ (i.e., high noise in measurement apparatus), repeated measurements lead to a near-maximal ($\approx 1$) fidelity of the post measured state (c.f.~\cite{Oreshkov2005}). 

\subsection{Noisy initial states}
\label{subsec:noisy_states}

We now consider the situation where the qubits in the unit states are sent through identical uncorrelated Pauli channels~\cite{nielsen2010}, resulting in mixed initial unit states. To probe this, let us assume the unit states $\ket{\psi_i}$ to take the form $\ket{\psi_i}=\cos\theta_i\ket{0_{B_i}}\ket{0_{A_i}}+\sin\theta_i\ket{1_{B_i}}\ket{1_{A_i}}$, which is a local rotation on $A$ qubits away from Eq.~(\ref{eq:ini_weak_states}). First, we consider the case of the phase-flip (PF) channels on all qubits $\alpha$, denoted by the map 
\begin{equation}\label{eq:PF_channel}
    \rho\to \Lambda^{\text{PF}}_\alpha(\rho)=\left(1-\frac{q}{2}\right)\rho+\frac{q}{2}\sigma^z\rho\sigma^z,
\end{equation}
where $\rho$ is a single qubit state and $0\leq q\leq 1$ is the noise strength with $q=0$ ($q=1$) corresponding to the absence of noise (maximum noise). To address entanglement over a bipartition of mixed states of arbitrary dimensions,  we employ negativity~\cite{vidal2002,peres1996,horodecki1996,zyczkowski1998,leggio2020} (see Appendix~\ref{app:PF_derivation} for definition). The $A:B$ bipartite negativity of unit state $\rho_i=\Lambda^{\text{PF}}_i(\ket{\psi_i})$ undergoing uncorrelated PF channel can be worked out to be (see Appendix.~\ref{app:PF_derivation}),
\begin{equation}\label{eq:Ei_PF}
    E_i^{\text{PF}}=(1-q)^2E_i,
\end{equation}
where $E_i$ is the bipartite negativity in the noiseless scenario ($q=0$). 

\begin{figure}
    \centering
    \includegraphics[width=0.8\linewidth]{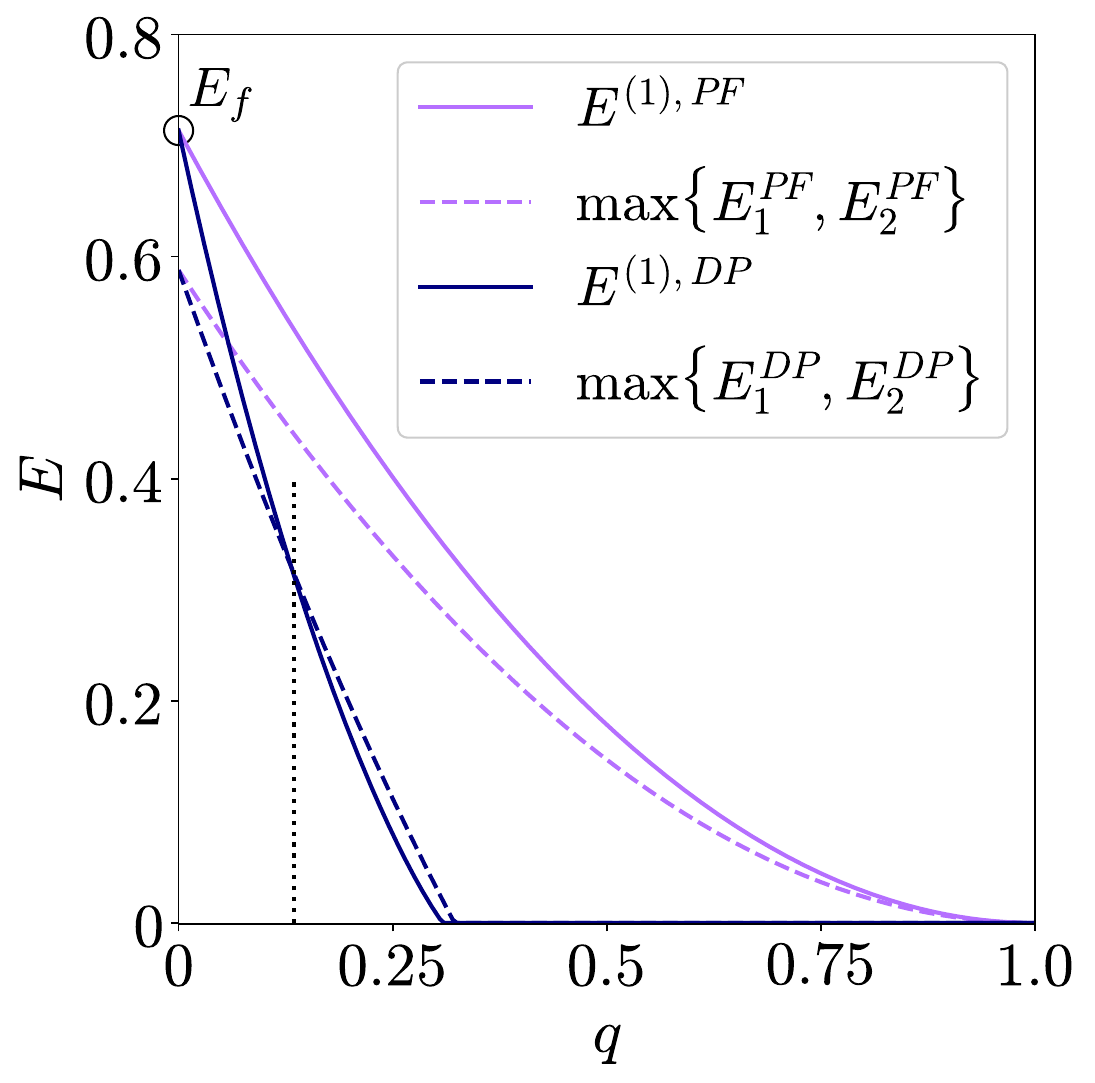}
    \caption{Variations of bipartite negativity $E^{(1)}$ (solid lines) and $\max\{E_1,E_2\}$ (dashed lines) in the cases of unit states undergoing uncorrelated phase-flip and depolarization channels. In the case of phase-flip channel, $E^{(1),\text{PF}}\geq\max\{E_1^{\text{PF}},E_2^{\text{PF}}\}$ in the entire range of $q$, while for the depolarization channel, this is possible only below a threshold around $q\approx0.135$. Here, for the unit states, we have fixed $\theta_1=\pi/12,\theta_2=\pi/10$ for demonstration.}
    \label{fig:noise_channels}
\end{figure}

Next, we perform a single round of two-qubit measurement $\mathcal{M}^{(1)}$ on the $B$ qubits of two mixed unit states $\rho_i\otimes\rho_j$. Similar to the case of the pure state (see Sec.~\ref{subsec:two_qubit_scheme}), the local unitary connection between the post-measured states corresponding to any two different outcomes, say $k=0$ and $k=1$, persists (see Appendix.~\ref{app:PF_derivation} for details), given by 
\begin{equation}\label{eq:noisy_LU_connection}
    \varrho_1^{(1)}=\left(U_A\otimes U_B\right) \varrho_1^{(0)} \left(U_A\otimes U_B\right).
\end{equation}
Post the first round of measurement $\mathcal{M}^{(1)}$,  the bipartite negativity of $\varrho_1^{(0)}$ turns out to be
\begin{equation}\label{eq:Ef_PF}
    E^{(1),\text{PF}}_0=(1-q)^2E^{(0)},
\end{equation}
where $E^{(0)}=\sqrt{E_i^2+E_j^2-E_i^2E_j^2}$ is the same corresponding to noiseless scenario (see Sec.~\ref{subsec:achieving_PCE}). Eqs.~(\ref{eq:Ei_PF}) and (\ref{eq:Ef_PF}), combined with $E^{(0)} \geq \max \{E_i,E_j \}$ (see Sec.~\ref{subsec:achieving_PCE}). Eqs.~(\ref{eq:Ei_PF}), results in
\begin{equation}\label{eq:bound_noisy}
    E^{(0),\text{PF}} \geq \max \{E_i^{\text{PF}},E_j^{\text{PF}} \},
\end{equation}
implying PCE in the first round of measurement irrespective of the value of $q$ (see Fig.~\ref{fig:noise_channels}). 

While analytically probing the cases of bit-flip, or depolarizing channels~\cite{nielsen2010} become cumbersome, our numerical investigations reveal the persistence of the local unitary equivalence of the post-measured states. However, in contrast to the case of the phase-flip channel, in the cases of the depolarizing noise, there exists a noise strength of the channel below (above) which bound equivalent to the one given in Eq.~(\ref{eq:bound_noisy}) for the corresponding channel holds (violates), and the noise strength depends on the input pure state. See Fig.~\ref{fig:noise_channels} for a demonstration in the case of the depolarizing channel. 

\noindent\textbf{Note 5.} It is worthwhile to point out that a similar variation of bipartite negativity under the phase flip channel of generalized GHZ as well as generalized W states are shown in~\cite{Jithin2023}, implying possible validity of Eq.~(\ref{eq:bound_noisy}) when unit states in the form of GHZ and W states under phase-flip noise are used as inputs to the protocol (see also Note 3).

\section{Application: Generating generalized GHZ states of arbitrary size}
\label{sec:GHZ_preperation}

As a specific application of the prescription for network growth with resultant PCE discussed in Secs.~\ref{sec:N-qubit_measurement} and~\ref{sec:repeated_two_qubit_measurement}, in this Section, we present the creation of generalized GHZ (gGHZ) states~\cite{greenberger1989} of $N$ qubits, given by
\begin{eqnarray}\label{eq:xim}
    \ket{\xi_{m}}&=&\cos \theta_{m}\ket{0_{B}}\ket{\mathbf{0}_A^m}+\sin\theta_{m}\ket{1_{B}}\ket{\mathbf{1}_A^m},
\end{eqnarray}
with $\ket{\mathbf{0}_A^m}=\bigotimes_{j=1}^m\ket{0_{A_{j}}}$, $\ket{\mathbf{1}_A^m}=\bigotimes_{j=1}^m\ket{1_{A_{j}}}$, $\theta_m\in[0,\pi/4]$ and $N=m+1$, starting from two-qubit ($m=1$) states of the same form. The following proposition is crucial for this. 

\proposition{6} \emph{Using the above prescription, a two-qubit state $\ket{\xi_{1}}$ can be merged with an $(m+1)$-qubit gGHZ state of the form $\ket{\xi_{m}}$ to form a $(m+2)$-qubit gGHZ state via a unitary operation $\mathcal{U}_A$ local to $A$.}

\begin{proof} 
Consider the following four steps of operations on the states $\ket{\xi_1}$ and $\ket{\xi_m}$.

\noindent\textbf{(a)} First, see that the two-qubit measurements on the $B$-qubits of $\ket{\xi_1}$ and $\ket{\xi_m}$ leads to the post measured states $\ket{\Phi_0}$ and $\ket{\Phi_1}$ with  
\begin{eqnarray}
    \ket{\Phi_0}&=&\ket{0_B}\big[\cos\theta_{1}\cos\theta_{m}\otimes_{j=1}^{m+1}\ket{0_{A_j}}\nonumber\\
    &&+\sin\theta_{1}\sin\theta_{m}\otimes_{j=1}^{m+1}\ket{1_{A_j}}\big]\nonumber\\
    &&+\ket{1_B}\big[\cos\theta_{1}\sin\theta_{m}\otimes_{j=1}^{m}\ket{0_{A_{1}}}\ket{1_{A_j}}\nonumber\\
    &&+\sin\theta_{1}\cos\theta_{m}\ket{1_{A_{1}}}\otimes_{j=1}^{m}\ket{0_{A_j}}\big],\nonumber\\
    \ket{\Phi_1}&=&\left[\prod_{i=2}^{m+1}\sigma^z_{A_i}\right]\ket{\Phi_0},
\end{eqnarray}
where we have labeled the $A-$qubit belonging to the state $\ket{\xi_1}$ as $A_1$ and $A_i\in\ket{\xi_m}$ as $A_2,A_3\cdots A_{m+1}$ for efficiently keeping tab on the $A$-qubits.

\noindent\textbf{(b)} Next, perform a series of controlled X operators, defined by 
\begin{eqnarray}
U^{\text{cx}}_{a_1,a_2}=\frac{1}{2}\left[(I+\sigma^z)_{a_1}I_{a_2}+(I-\sigma^z)_{a_1}\sigma^x_{a_2}\right]
\end{eqnarray}
for qubit-pairs $(a_1,a_2)$, on the \emph{nearest-neighbor}  pairs of $A$-qubits in $\ket{\Phi_0}$. The \emph{overall} $A$-unitary is defined by 
\begin{eqnarray}
U^{\text{cx}}_{A}=\left\{\begin{array}{l}
U^{\text{cx}}_{A_1,A_2}\text{ for }m=1,\\
U^{\text{cx}}_{A_1,A_2}V_A^m\text{ for }m>1, 
\end{array}\right.
\label{eq:general_CXP} 
\end{eqnarray}
with 
\begin{eqnarray}
    V_A^m=\prod_{i=0}^{m-2} U^{\text{cx}}_{A_{m-i+1},A_{m-i}} U^{\text{cx}}_{A_{m-i},A_{m-i+1}},
    \label{eq:V}
\end{eqnarray}
leading to 
\begin{eqnarray}
U^{\text{cx}}_{A}\ket{\Phi_0}&=&\ket{0_{B}}\ket{\alpha_{A_1}}\bigotimes_{j=2}^{m+1}\ket{0_{A_j}}+\ket{1_{B}}\ket{\beta_{A_1}}\bigotimes_{j=2}^{m+1}\ket{1_{A_j}},\nonumber\\ \label{eq:post_meas_after_CXP}    
\end{eqnarray} 
where we have defined 
\begin{eqnarray}
    \ket{\alpha_{A_1}}&=&\cos\theta_{1}\cos\theta_{m}\ket{0_{A_1}}+\sin\theta_{1}\sin\theta_{m}\ket{1_{A_1}},\nonumber\\
    \ket{\beta_{A_1}}&=&\cos\theta_{1}\sin\theta_{m}\ket{0_{A_1}}+\sin\theta_{1}\cos\theta_{m}\ket{1_{A_1}}.
\end{eqnarray}
Note here that the ordering of control and target qubits arranged in Eqs.~(\ref{eq:general_CXP}) and (\ref{eq:V}) is important to arrive at the desired state. Also the unitary operator $V_A^m$ starts operating only when $m>1$.  

\noindent\textbf{(c)} We now define 
\begin{equation}
\cos\theta_{m+1}=\sqrt{\cos^2\theta_{m}\cos^2\theta_{1}+\sin^2\theta_{m}\sin^2\theta_{1}},
\label{eq:theta_final_single_iteration}
\end{equation}
and perform a controlled rotation 
\begin{eqnarray}
    R_{A_2,A_1}=\frac{1}{2}[I_{A_1}\otimes(I+\sigma_z)_{A_2} + J_{A_1}\otimes(I-\sigma_z)_{A_2}]
\end{eqnarray}
over the qubit-pair $(A_1,A_2)$, with 
\begin{eqnarray}
    J_{A_1}&=&\frac{\sin2\theta_{1}\cos2\theta_{m}}{\sin2\theta_{m+1}}\sigma^z-\frac{\sin2\theta_{m}}{\sin2\theta_{m+1}}\sigma^x,
\end{eqnarray}
where $A_2$ ($A_1$) is the control (target) qubit

\noindent\textbf{(d)} The last step is a local rotation on the qubit $A_1$, given by 
\begin{eqnarray}
L_{A_1}&=&-\frac{\cos\theta_{1}\cos\theta_{m}}{\cos\theta_{m+1}} I+\text{i}\frac{\sin\theta_{1}\sin\theta_{m}}{\cos\theta_{m+1}}\sigma^y.     
\end{eqnarray}
Applying $L_{A_1}R_{A_1,A_2}$ on $U_{A}^{\text{cx}}\ket{\Phi_0}$ results in $\ket{\xi_{m+1}}$ as 
\begin{eqnarray}\label{eq:single_step_gGHZ}
\ket{\xi_{m+1}}&=&L_{A_1}R_{A_1,A_2}U_{A}^{\text{cx}}\ket{\Phi_0}\nonumber\\
&=& \cos\theta_{m+1} \ket{0_B}\bigotimes_{j=1}^{m+1}\ket{0_{A_j}}\nonumber\\&&+\sin\theta_{m+1} \ket{1_B}\bigotimes_{j=1}^{m+1}\ket{0_{A_j}}.
\end{eqnarray}
Note that the overall unitary
\begin{eqnarray}
\mathcal{U}_A=L_{A_1}R_{A_2,A_1}U^{\text{cx}}_{A} 
\label{eq:A-unitary}
\end{eqnarray} 
is local to the $A$-qubits. Hence the proof. 
\end{proof}

\begin{figure*}
    \centering
    \includegraphics[width=\linewidth]{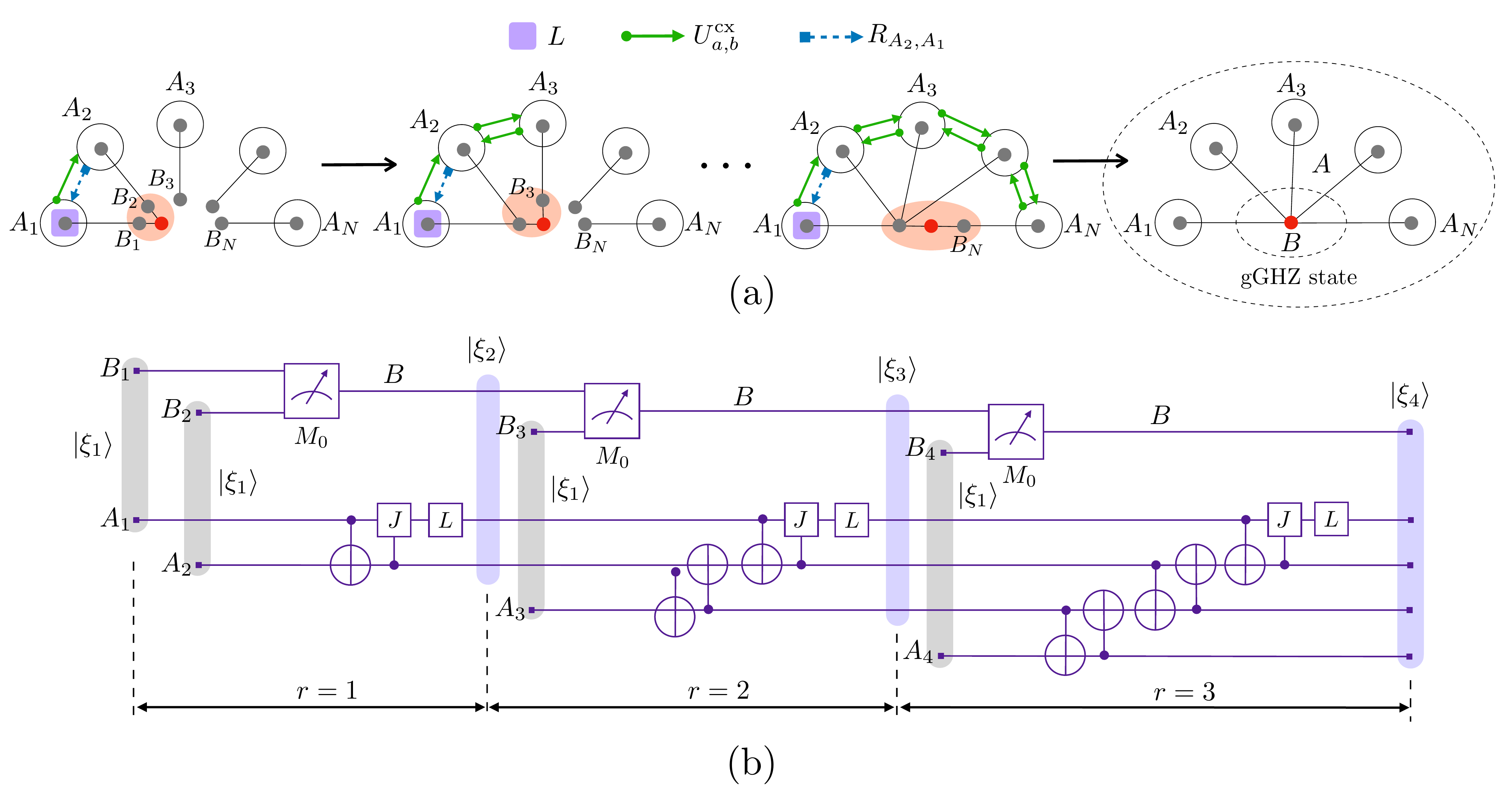}
    \caption{(a) Protocol for deterministically creating large gGHZ states, starting from multiple copies of two-qubit states $\ket{\xi_1}$, using the two-qubit repeated measurement scheme. The controlled unitary operations $U_{a,b}^{\text{cx}}$ ($a,b\in\{A_i;i=1,2,\cdots\}$) and $R_{A_2,A_1}$ are demonstrated by arrows, which are from the \emph{control} to the \emph{target} qubits. (b) A quantum circuit for preparing a five-qubit gGHZ state out of four two-qubit non-maximally entangled states via three iterations of the protocol.}
    \label{fig:gGHZ}
\end{figure*}

The unitary operation $\mathcal{U}_A$ required to produce an $\ket{\xi_m}$ combining $\ket{\xi_{m-1}}$ and $\ket{\xi_{1}}$ contains $2(m-1)$ controlled operations, leading to $\mathcal{N}=\sum_{i=1}^m 2(i-1)=m^2-m$ as the number of controlled operations required to produce a state of the form $\ket{\xi_{m}}$ by joining $m$ $\ket{\xi_1}$'s. Since the two-qubit measurement can be performed as in Fig.~\ref{fig:two_qubit_measurement_circuit}, the number of two-qubit operations corresponding to the two-qubit measurements varies as $\sim m$, which is the sub-leading order in $m$ in $\mathcal{N}$.  Starting from a collection of two-qubit non-maximally entangled states of the form $\ket{\xi_1}$, the protocol for creating gGHZ states of arbitrary sizes exploiting Proposition 6 is outlined in Fig.~\ref{fig:gGHZ}(a), while the quantum circuit required for three rounds of the protocol is given in Fig.~\ref{fig:gGHZ}(b) (in Appendix~\ref{app:GHZ_preperation}, we provide a pseudo code for the outlined gGHZ state preparation protocol). Note that creating gGHZ states of arbitrary number of qubits require availability of a pool of adequate number of two-qubit non-maximally entangled states of the form $\ket{\xi_1}$. Also note that given propositions 4 and 6 it is possible to create an $N$-qubit GHZ state if at least one of the available two-qubit entangled states is a Bell state (i.e., $\theta=\pi/4$). This enables creation of maximally-entangled state over any two qubits shared between two parties even in the scarcity of  Bell states via (a) creating a GHZ state including all qubits using the proposed protocol, and (b) subsequently performing single-qubit measurements in $\sigma^x$ basis on the remaining qubits to create a maximally-entangled state among the desired qubits~\cite{verstraete2004,verstraete2004a,popp2005}.

\section{Conclusions and outlook}
\label{sec:conclusion}

Growing quantum networks and entanglement concentration are important ideas in the light of practical quantum communication. In this paper, we devise a graph states-assisted $N$-qubit measurement scheme that truncates the Hilbert space of $N$ qubits to that of a single qubit. This measurement, when applied to $N$ entangled qubit-pairs, results in equiprobable local unitary-connected post-measured states, making the measurement effectively deterministic as long as the entanglement of the generated states is concerned. We prove that the measurement results in a profitable concentration of entanglement, where each of the post-measured states possesses higher bipartite entanglement compared to the same for any of the individual entangled qubit-pairs participating in the measurement. Further, we propose a a scalable and easy-to-deploy repetitive two-qubit measurement protocol as an alternative to the joint $N$-qubit measurement scheme, and analytically determine the lower and the upper bounds of the concentrated entanglement. We also analyse the robustness of the protocol against various imperfections that may arise in practical scenarios, including imperfections in components of the circuit implementing the two-qubit measurement and noise in the input two-qubit states. We also show that the scheme can be used to construct highly entangled multi-qubit generalized GHZ states out of weakly-entangled two-qubit unit states. 

We point out here that unlike~\cite{Gu2006,Zhao2013}, our scheme is free from any restriction on the entanglement content of the available non-maximal entangled qubit-pairs, and requires no additional auxiliary system, as opposed to the circuit-based prescription presented in~\cite{Gu2006}. With the existence of probabilistic protocols~\cite{W_Tashima2009,W_Zang2015,W_Sharma2020} to grow $W$ states~\cite{dur2000}, our work poses the immediate question as to whether a measurement-based yet deterministic preparation of large $W$ state, and more generally Dicke states~\cite{dicke1954}  is possible. Also, the question of extending the devised scheme towards entanglement percolation~\cite{Acin2007} is underlined by the application, although not towards a deterministic scheme, of the use of two-qubit measurement-based schemes in related problems~\cite{Perseguers2010,Meng2021}.

\acknowledgements 
H.K.J. thanks the Prime Minister's Research Fellowship (PMRF) program, Government of India, for the financial support. A.K.P acknowledges the support from the Anusandhan National Research Foundation (ANRF) of the Department of Science and Technology (DST), India, through the Core Research Grant (CRG) (File No. CRG/2023/001217, Sanction Date 16 May 2024).

\onecolumngrid 
\appendix

\section{Two-qubit measurement}
\label{app:two_qubit_measurement}

In this Section, we discuss the two-qubit measurement considered in the repetitive two-qubit measurement protocol, which corresponds to the special case of $N=2$ in the $N$-qubit measurement of Eq.~(\ref{eq:star_twoqubit_meas}). The two qubit Hadamard operator $H_{B_i}\otimes H_{B_j}$ being self-inverse, its columns can be identified as $\ket{e_b}$ $(\ket{b})$, for $b=00,01,10,11$, when written in the $\{\ket{e_b}\}$ $(\{\ket{b}\})$ basis. Also, we assume that the two-qubit measurements are performed on two arbitrary pairs with $A:B$ bipartite entanglement $E_i$ and $E_j$ and that of the post measured state is $E_f$. There are only three distinct possible choices of the zero'th pair $(b_0,b_0^\prime)$ in this case given by $(00,01),(00,10)$ and $(00,11)$. Corresponding to each of the choices, the Schmidt coefficients of the post-measured states (Eq.~(\ref{eq:cos_thetaf_supple})) can be obtained as
\begin{eqnarray}
    (00,01) &\rightarrow& {\cos\theta_{f}} = \cos\theta_j,\\
    (00,10) &\rightarrow&{\cos\theta_{f}} = \cos\theta_i,\\
    (00,11) &\rightarrow&{\cos\theta_{f}} = \sqrt{\cos^2\theta_i\cos^2\theta_j+\sin^2\theta_i\sin^2\theta_j},\nonumber\\
\end{eqnarray}
leading to their bipartite entanglements, 
\begin{eqnarray}
    (00,01) &\rightarrow& E_f=E_j,\\
    (00,10) &\rightarrow& E_f=E_i,\\
    (00,11) &\rightarrow&E_f=\sqrt{E_i^2 + E_j^2-E_i^2 E_j^2}\label{eq:Ef_formula}.  
\end{eqnarray}
Note that while the choice of the zero'th pair $(00,01)$ and $(00,10)$ provides $E_f$ to be equal to the entanglement in an already existing pair ($E_j$ or $E_i$ respectively), to achieve $E_f>\max\{E_i,E_j\}$, one needs to choose the zero'th pair to be $(00,11)$. This can be shown in the same fashion as in the proof of proposition 4. The measurement corresponding to this choice of zero'th pair can be performed with the quantum circuit given in Fig.~\ref{fig:two_qubit_measurement_circuit} of the main text.

\section{Merging two-qubit unit states}
\label{app:two_qubit_unit_states}

We apply the repetitive two-qubit measurement protocol over {identical} two-qubit states, $\ket{\psi_0}$ with entanglement $E_0$ (corresponding to $\theta_i=\theta_0\forall i$), to generate a large bipartite-entangled states on multiple qubits. Post $r$ rounds of measurements merging $(r+1)$ two-qubit states the post measurement state (Eq.~(\ref{eq:post_measured_state_r_rounds})) takes the form,
\begin{eqnarray}
    \ket{\Phi_0^{(r)}}&=&\ket{0}_B\sum_{k=0,2,\cdot\cdot\cdot}^{r+1} (\cos\theta_0)^{r+1-k} (\sin\theta_0)^{k}\mathcal{P}(\ket{0}^{\otimes r+1-k}\ket{1}^{\otimes k})+\ket{1}_B\sum_{k=1,3,\cdot\cdot\cdot}^{r+1} (\cos\theta_0)^{r+1-k} (\sin\theta_0)^{k}\mathcal{P}(\ket{0}^{\otimes r+1-k}\ket{1}^{\otimes k}),\nonumber\\
\end{eqnarray}    
where $\mathcal{P}(.)$ represents all possible permutations of $r+1-k$ qubits in state $\ket{0}$ and $k$ qubits in $\ket{1}$, suggesting an invariance of $\ket{\Phi_0^{(r)}}$ with respect to all possible permutations of $A_i$s, leading to identical reduced density matrices of $B$ and any of the $A_i$ qubit,
\begin{eqnarray}
    \rho_{BA_i}^{(r)}=\text{Tr}_{\{A_j;j\neq i\}} \ket{\Phi_0^{(r)}}\bra{\Phi_0^{(r)}}.
\end{eqnarray}
Explicitly, one can calculate,  
\begin{eqnarray}
    \rho_{BA_i}^{(r)}=\frac{1}{2}\begin{bmatrix}
        a_{00} & 0 & 0 & \sqrt{a_{00}a_{11}}\\
        0 & a_{01} & \sqrt{a_{01}a_{10}} & 0\\
        0 & \sqrt{a_{01}a_{10}} & a_{10} & 0\\
        \sqrt{a_{00}a_{11}} & 0 & 0 & a_{11}\\        
    \end{bmatrix},
\end{eqnarray}
with 
\begin{eqnarray}
    a_{00}&=&\cos^2\theta_0\left[1+(\cos2\theta_0)^r\right],\nonumber \\
    a_{01}&=&\sin^2\theta_0\left[1-(\cos2\theta_0)^r\right],\nonumber \\
    a_{10}&=&\cos^2\theta_0\left[1-(\cos2\theta_0)^r\right],\nonumber \\
    a_{11}&=&\cos^2\theta_0\left[1+(\cos2\theta_0)^r\right].
\end{eqnarray}
In the case of a mixed two-qubit state concurrence is given by $E(\rho)=\max\{0,\lambda_1-\lambda_2-\lambda_3-\lambda_4\}$ where $\lambda_i$'s are the eigenvalues of the matrix $\sqrt{\rho \tilde\rho }$ arranged in the decreasing order. Here, $\tilde{\rho}=\sigma^y_1\sigma^y_2\rho^*\sigma^y_1\sigma^y_2$ where $\rho^*$ being the complex conjugate of $\rho$. Thus, the concurrence between $B$ and $A_i$ after the $r$th round of measurement can be  evaluated as,
\begin{eqnarray}
    E_{BA_i}^{(r)}=E_0(1-E_0^2)^{r/2}.
\end{eqnarray}
It is evident that with increasing $r$, $E^{(r)}_{BA_i}$ decreases monotonically, and eventually vanishes. Further, $E^{(r)}_{A_iA_j}=0$ for any $(A_i,A_j)$ pair with $j\neq i$, and $A_i,A_j\in A$ as the two-qubit measurements are local to the subsystem $B$. The monogamy score of concurrence squared~\cite{Dhar2017,*Zong2022} (cf.~\cite{Coffman2000}) corresponding to the qubit $B$ as the nodal observer after $r$ rounds of two-qubit measurements is defined as $\delta_B^{(r)}=(E^{(r)})^2-(r+1)(E_{BA_i}^{(r)})^2$ and it turns out to be,  
\begin{eqnarray}
    \delta_B^{(r)}&=&1-(1-E_0^2)^{r}(1+rE_0^2),
\end{eqnarray}
which is a monotonically increasing function of $r$ for fixed $E_0$.

Aware of the fact that the generated states are multi-qubit states, we further investigate the multipartite quantum correlations present in the generated states, as quantified by the generalized geometric measure (GGM)~\cite{sende2010a,*sende2012} (cf.~\cite{SHIMONY1995,*Barnum_2001,*Wei2003}). GGM of a pure state $\ket{\Phi}$ is given by $\text{GGM}(\ket{\Phi})=1-\max\{\lambda_i^2\}$ where $\lambda_i$ is the maximal Schmidt coefficient in a bipartition of $\ket{\Phi}$ and the maximization in the definition is taken over all possible bipartitions of the state. Due to proposition 4, the maximum Schmidt coefficient in the $B:A_{(r)}$ partition decreases monotonically with $r$, implying that the GGM is not decided by this partition. Our numerical analysis suggests that the maximum of the Schmidt coefficients among all possible bipartitions of the system $BA_1A_2\cdots A_{r+1}$ is always provided by the bipartition $A_i:B\mathcal{A}$ with $\mathcal{A}=\bigcup_{j\neq i}A_i$, and $A_i$ is determined by the two-qubit system with the minimum entanglement, i.e., $\min\{E_1,E_2,\cdots,E_{r+1}\}$.

\section{Unit states under phase-flip channel}
\label{app:PF_derivation}

Here we present the details corresponding to the results discussed in Sec.~\ref{subsec:noisy_states}. Consider unit states $\ket{\psi_i}=\cos\theta_i\ket{0_{B_i}}\ket{0_{A_i}}+\sin\theta_i\ket{1_{B_i}}\ket{1_{A_i}}$ undergoing the phase-flip channel (Eq.~(\ref{eq:PF_channel})) as 
\begin{eqnarray}
    \rho_i&=&\Lambda^{\text{PF}}(\ket{\psi_i})\nonumber\\
    &=&(1-p)\ket{\psi_i}\bra{\psi_i}+p\ket{\tilde\psi_i}\bra{\tilde\psi_i},\label{eq:PF_mixed_unit_state_app}
\end{eqnarray}
where $p=q\left(1-\frac{q}{2}\right)$ and $\ket{\tilde\psi_i}=\cos\theta_i\ket{0_{B_i}}\ket{0_{A_i}}-\sin\theta_i\ket{1_{B_i}}\ket{1_{A_i}}$. The negativity~\cite{vidal2002,peres1996,horodecki1996,zyczkowski1998,leggio2020} of a bipartite mixed state $\rho_{A:B}$ is given by 
\begin{equation}
    E(\rho_{A:B})=2\sum_{\lambda_i<0}|\lambda_i|,
\end{equation}
with $\lambda_i$'s being the eigenvalues of the matrix $\rho^{T_B}$ obtained by performing a partial transpose of $\rho$ with respect to the subsystem $B$. Note that the factor two is added to normalize the value of negativity to be within zero and one. Therefore partial transposing $\rho_i$,
\begin{equation}
    \rho_i^{T_{B_i}}=\begin{bmatrix}
        \cos^2\theta_1 & 0 & 0 & 0 \\
        0 & 0 & [(1-2p)\sin2\theta_1]/2 & 0 \\
        0 & [(1-2p)\sin2\theta_1]/2 & 0 & 0 \\
        0 & 0 & 0 & \sin^2\theta_1         
    \end{bmatrix},
\end{equation}
leads to the negativity of a unit state under PF channel to be,
\begin{equation}\label{eq:Ei_PF_app}
    E_i^{\text{PF}}=(1-q)^2E_i,
\end{equation}
where $E_i=\sin2\theta_i$ is the negativity when $q=0$.
A single round of two-qubit measurement $\mathcal{M}^{(1)}$ (see Sec.~\ref{sec:repeated_two_qubit_measurement}) on the $B$ qubits of two such mixed unit states $\rho_i\otimes\rho_j$ results in a post measured mixed state constituted of four components mixed with appropriate probabilities. The component wise output of the circuit in Fig.~\ref{fig:two_qubit_measurement_circuit} excluding the final $\sigma^z$ measurement of $B_j$ qubit is as follows.
\begin{eqnarray}
    \ket{\psi_i}\ket{\psi_j}&\xrightarrow{(1-p)^2}&\frac{\ket{\Phi^{(1)}_0}\ket{0_{B_j}}+\ket{\Phi^{(1)}_1}\ket{1_{B_j}}}{\sqrt{2}},\nonumber\\
    \ket{\tilde\psi_i}\ket{\psi_j}&\xrightarrow{p(1-p)}&\sigma^z_{A_1}\frac{\ket{\Phi^{(1)}_0}\ket{0_{B_j}}+\ket{\Phi^{(1)}_1}\ket{1_{B_j}}}{\sqrt{2}},\nonumber\\
    \ket{\psi_i}\ket{\tilde\psi_j}&\xrightarrow{p(1-p)}&\sigma^z_{A_2}\frac{\ket{\Phi^{(1)}_0}\ket{0_{B_j}}+\ket{\Phi^{(1)}_1}\ket{1_{B_j}}}{\sqrt{2}},\nonumber\\
    \ket{\tilde\psi_i}\ket{\tilde\psi_j}&\xrightarrow{~~~p^2~~~}&\sigma^z_{A_1}\sigma^z_{A_2}\frac{\ket{\Phi^{(1)}_0}\ket{0_{B_j}}+\ket{\Phi^{(1)}_1}\ket{1_{B_j}}}{\sqrt{2}}.        
\end{eqnarray}
\normalsize
where the probabilities of occurrence of each state are indicated above the arrows,  and
\begin{eqnarray}
    \ket{\Phi^{(1)}_0}&=&[\cos\theta_i\cos\theta_j\ket{0_{A_i}0_{A_j}} + \sin\theta_i\sin\theta_j\ket{1_{A_i}1_{A_j}}]\ket{0_{B_i}}+ [\cos\theta_i\sin\theta_j\ket{0_{A_i}1_{A_j}} + \sin\theta_i\cos\theta_j\ket{1_{A_i}0_{A_j}}]\ket{1_{B_i}},\nonumber\\
    \ket{\Phi^{(1)}_1}&=&\sigma^z_{A_2}\ket{\Phi^{(1)}_0}.
\end{eqnarray}
Therefore, the final $\sigma^z$ measurement of $B_j$ qubit results in a local unitary connected mixed states, 
\begin{equation}
    \varrho^{(1)}_1=\sigma^z_{A_2}\varrho^{(1)}_0\sigma^z_{A_2}
\end{equation}
where the normalized post measured state $\varrho^{(1)}_0$ is given by 
\begin{eqnarray}
    \varrho^{(1)}_0&=&(1-p)^2\ket{\Phi^{(1)}_0}\bra{\Phi^{(1)}_0}+p^2\sigma^z_{A_iA_j}\ket{\Phi^{(1)}_0}\bra{\Phi^{(1)}_0}\sigma^z_{A_iA_j}+p(1-p)[\sigma^z_{A_i}\ket{\Phi^{(1)}_0}\bra{\Phi^{(1)}_0}\sigma^z_{A_i}+\sigma^z_{A_j}\ket{\Phi^{(1)}_0}\bra{\Phi^{(1)}_0}\sigma^z_{A_j}].
\end{eqnarray}    

In order to evaluate the negativity over the partition $A:B$  of $\varrho^{(1)}_0$, we perform the partial transposition with respect to the $B_i$ qubit, i.e.,  $(\varrho^{(1)}_0)^{T_{B_i}}$, followed by a rearrangement of basis elements of $B_iA_iA_j$ in the order $\{\ket{000},\ket{011},\ket{101},\ket{110},\ket{001},\ket{010},\ket{100},\ket{111}\}$,  yielding a block diagonal structure for $(\varrho^{(1)}_0)^{T_{B_i}}$. While the first two $2\times 2$ blocks having only positive eigenvalues do not contributes to the negativity,  the third $4\times 4$ block has the structure
\begin{equation}
    (\varrho^{(1)}_0)^{T_{B_i}}_{4\times 4}=\begin{bmatrix}
        0 & R\\
        R^T & 0
    \end{bmatrix},
\end{equation}
where the $2\times 2$ blocks 
\begin{equation}
    R=(1-2p)\begin{bmatrix}
        \cos^2\theta_1\cos\theta_2\sin\theta_2 & \cos\theta_1\sin\theta_1\sin^2\theta_2 \\
        \cos\theta_1\sin\theta_1\cos^2\theta_2 & \sin^2\theta_1\cos\theta_2\sin\theta_2
    \end{bmatrix}.
\end{equation}
Diagonalizing $(\varrho^{(1)}_0)^{T_{B_i}}_{4\times 4}$ yields the non-trivial eigenvalues $\pm(1-q)^2E^{(0)}/2$ with $E^{(0)}=\sqrt{E_i^2+E_j^2-E_i^2E_j^2}$ (Eq.~(\ref{eq:Ef_formula})) being the bipartite negativity for $q=0$. Therefore, negativity of $\varrho_0^{(1)}$ over the bipartition $A:B$ is given by
\begin{equation}\label{eq:Ef_PF_app}
    E^{(0),\text{PF}}=(1-q)^2E^{(0)}.
\end{equation}
Note that with $E^{(0)}\geq \max\{E_i,E_j\}$, Eqs.~(\ref{eq:Ei_PF}) and (\ref{eq:Ef_PF}) ensures 
\begin{equation}
    E^{(0),\text{PF}}\geq \max\{E_i^\text{PF},E_j^\text{PF}\}.
\end{equation}

\section{Pseudocode for the Generalized GHZ state preparation}
\label{app:GHZ_preperation}

Here, we present a pseudocode representing the proposal for generating gGHZ states of arbitrary sizes. The explicit forms of the unitary operators $L_{A_1}$, $R_{A_2,A_1}$, and $U_{A_1,A_2}^{\text{cx}}$ are given in the Section \ref{sec:GHZ_preperation} of the main text, while the code is represented in the form of a quantum circuit in Fig.~\ref{fig:gGHZ}(b).

\vspace{0.5cm}

\noindent\textbf{Preparation of $N$-qubit gGHZ state.}

\vspace{0.25cm}

\noindent\textbf{input:} $N-1$ qubit-pairs in states $\ket{\xi_1}$ ($m=1$ in Eq.~(\textcolor{red}{16}) of main text). 

\noindent\textbf{steps:}
\begin{enumerate}
    \item[\textbf{1.}] Choose two states $\ket{\xi_1^{(1)}}$ and $\ket{\xi_1^{(2)}}$ of the form $\ket{\xi_1}$ ($m=1$ in Eq.~(\textcolor{red}{16})), defined on the qubit-pairs $A_1B_1$ and $A_2B_2$. 
    \item[\textbf{2.}] Perform the following operations maintaining the order to create a three-qubit bipartite state $\ket{\xi_2}$, where the partition $B$ contains one qubit, and the partition $A$ contains qubits $A_1$ and $A_2$. 
    \begin{enumerate}
        \item[\textbf{a.}] A two-qubit measurement (see Eq.~(\textcolor{red}{9}) of the main text) on the qubits $B_1$ and $B_2$.
        \item[\textbf{b.}] A controlled $X$ operation $U^{\text{cx}}_{A_1,A_2}$. 
        \item[\textbf{c.}] A controlled rotation $R_{A_2,A_1}$.
        \item[\textbf{d.}] A single-qubit rotation $L_{A_1}$.
    \end{enumerate}
    \item[\textbf{3.}] Merge another two-qubit state $\ket{\xi_1^{(3)}}$ with the three-qubit state $\ket{\xi_2}$ with to create a four-qubit state $\ket{\xi_3}$ via the following operations. 
    \begin{enumerate}
        \item[\textbf{a.}] Perform a two-qubit measurement on the $B$-qubits of $\ket{\xi_2}$ and $\ket{\xi_1^{(3)}}$ (Eq.~(\textcolor{red}{9}) of the main text).
        \item[\textbf{b.}] Perform the controlled $X$ operations $U^{\text{cx}}_{A}$ on the $A$-qubits (Eq.~(\textcolor{red}{20}) of the main text).
        \item[\textbf{c.}] The controlled rotation $R_{A_2,A_1}$ from $A_2$ to $A_1$ (Eq.~(\textcolor{red}{18}) of the main text).
        \item[\textbf{d.}] The single-qubit rotation $L_{A_1}$ (Eq.~(\textcolor{red}{17}) of the main text).
    \end{enumerate}
    \item[\textbf{4.}] Continue \textbf{step 3} $(N-3)$-times, where the $k$th execution of \textbf{step 3} merges a two-qubit state $\ket{\xi_1}$ with a $(k+2)$-qubit state $\ket{\xi_{k+1}}$ to generate a $(k+3)$-qubit state $\ket{\xi_{k+2}}$. In each step perform the two qubit measurement followed by the unitary operation $U^{\text{cx}}_{A}R_{A_2,A_1}L_{A_1}$.
\end{enumerate}
\noindent\textbf{output:} An $N$-qubit gGHZ state $\ket{\xi_{N-1}}$.

\twocolumngrid 

\bibliography{ref}

\end{document}